\documentclass{article}
\pdfoutput=1
\usepackage{amsmath,amssymb,graphicx}
\usepackage{hyperref}
\usepackage[T1]{fontenc}
\usepackage[utopia]{mathdesign}
\usepackage[font=small,labelfont=bf]{caption}
\usepackage{longtable}
\usepackage{verbatim}
\usepackage{amsfonts}
\usepackage{cite}

\usepackage[usenames,dvipsnames]{xcolor}
\definecolor{burgundy}{rgb}{0.565,0.0,0.125}

\usepackage{afterpage}

\newcommand{\bear}{\begin{array}}
\newcommand{\ear}{\end{array}}

\newcommand{\beq}{\begin{eqnarray}}
\newcommand{\eeq}{\end{eqnarray}}
\newcommand{\beqa}{\begin{eqnarray}}
\newcommand{\eeqa}{\end{eqnarray}}

\def\OMIT#1{{}}
\newcommand{\lsim}{\mathrel{\rlap{\lower4pt\hbox{\hskip1pt$\sim$}}
    \raise1pt\hbox{$<$}}}         
\newcommand{\gsim}{\mathrel{\rlap{\lower4pt\hbox{\hskip1pt$\sim$}}
    \raise1pt\hbox{$>$}}}         

\newcommand{\WW}{\mathcal{W}}

\usepackage{titlesec}

\titleformat{\section}
{\color{burgundy}\normalfont\Large\bfseries}
{\color{burgundy}\thesection}{1em}{}
\titleformat{\subsection}
{\color{burgundy}\normalfont\large\bfseries}
{\color{burgundy}\thesubsection}{1em}{}

\title{\bf \color{burgundy} In Wino Veritas?\\
{\Large Indirect Searches Shed Light on Neutralino Dark Matter }}

\author{JiJi Fan and Matthew Reece\\
{\em Department of Physics, Harvard University, Cambridge, MA 02138, USA}}

\begin{document}
\maketitle

\begin{abstract}
Indirect detection constraints on gamma rays (both continuum and lines) have set strong constraints on wino dark matter. By combining results from Fermi-LAT and HESS, we show that: light nonthermal wino dark matter is {\em strongly} excluded; thermal wino dark matter is allowed only if the Milky Way dark matter distribution has a significant ($\gsim 0.4$ kpc) core; and for plausible NFW and Einasto distributions the entire range of wino masses from 100 GeV up to 3 TeV can be excluded. The case of light, nonthermal wino dark matter is particularly interesting in scenarios with decaying moduli that reheat the universe to a low temperature. Typically such models have been discussed for low reheating temperatures, not far above the BBN bound of a few MeV. We show that constraints on the allowed wino relic density push such models to higher reheating temperatures and hence heavier moduli. Even for a flattened halo model consisting of an NFW profile with constant-density core inside 1 kpc and a density near the sun of 0.3 GeV/cm$^3$, for 150 GeV winos current data constrains the reheat temperature to be above 1.4 GeV. As a result, for models in which the wino mass is a loop factor below $m_{3/2}$, the data favor moduli that are more than an order of magnitude heavier than $m_{3/2}$. We discuss some of the sobering implications of this result for the status of supersymmetry. We also comment on other neutralino dark matter scenarios, in particular the case of mixed bino/higgsino dark matter. We show that in this case, direct and indirect searches are complementary to each other and could potentially cover most of the parameter space. 
\end{abstract}


\section{Introduction}
Supersymmetry has long been a favorite theoretical candidate for physics beyond the Standard Model. Its phenomenological consequences include the possibility of natural electroweak symmetry breaking, gauge coupling unification, and weakly-interacting dark matter candidates stabilized by matter parity. Our goal in this paper is to explore the implications of indirect searches for neutralino dark matter in the MSSM using gamma rays, both continuum and lines. We are particularly interested in the case of light wino dark matter, and in the implications of constraints on its abundance for nonthermal cosmological histories. Such cosmologies are now motivated by LHC constraints on supersymmetry and insights from top-down models of SUSY breaking. We will also initiate a study of mixed states as neutralino dark matter candidates, in particular, a mixed bino/higgsino scenario, to illustrate that ongoing indirect searches of gamma rays could be a powerful complementary probe to direct dark matter searches.

The discovery of a 125 GeV Higgs boson at the LHC has thrown the naturalness argument for supersymmetry into sharp relief. The reason is that the MSSM requires large supersymmetry-breaking stop soft masses or $A$-terms to lift the Higgs mass above the $Z$ boson mass~\cite{Haber:1990aw,Barbieri:1990ja}. These large stop masses and $A$-terms feed quadratically into one-loop corrections to the soft mass $m_{H_u}^2$, directly making the theory more fine-tuned. As a result, models for supersymmetry at the weak scale have split into two possibilities with very different implications for naturalness and the fine-tuning of our universe. The first route, ``natural supersymmetry,'' invokes new contributions to the Higgs mass (generally arising at tree-level) to achieve $m_h \approx 125$ GeV while still keeping the stop soft masses and $A$-terms small, as required for low fine-tuning~\cite{Dimopoulos:1995mi,Cohen:1996vb}. It requires significant model-building in both the Higgs and flavor sectors, and is under increasing strain from direct searches at the LHC~\cite{Lodone:2012kp}. The second route is to postulate a spectrum of the ``mini-split'' type, with scalars somewhat (typically a loop factor) heavier than gauginos. This type of spectrum is predicted by both the simplest version of anomaly mediation~\cite{Giudice:1998xp, Randall:1998uk} and a wide variety of moduli mediation scenarios~\cite{Choi:2005ge,Choi:2005uz,Conlon:2006us,Conlon:2006wz,Acharya:2007rc,Acharya:2008zi}. Initially it was treated as an embarrassing prediction to be solved by working harder at model-building~\cite{Giudice:1998xp,Randall:1998uk}. However, its virtues include simplicity and amelioration of SUSY flavor and CP problems, so in the last decade it has begun to be considered as a viable possibility despite requiring fine-tuned EWSB, beginning with a prescient paper by James Wells~\cite{Wells:2003tf} and continuing with a variety of subsequent work on ``split SUSY''~\cite{ArkaniHamed:2004fb,Wells:2004di,Acharya:2007rc,Acharya:2008zi,Hall:2011jd,Arvanitaki:2012ps,ArkaniHamed:2012gw,Hall:2012zp}. This route gives up on strict naturalness; it allows supersymmetry to explain and stabilize {\em most} of the hierarchy between weak and Planck scales, but has significant residual fine-tuning.

For the second route, the most natural candidate for dark matter is wino dark matter. It could have a thermal history, if its mass is about 2.8 TeV~\cite{Hisano:2006nn, Cirelli:2007xd}.\footnote{We use the PLANCK + WP value, $\Omega_{{\rm DM}}h^2$ = $0.1199\pm0.0027$ at 68\% CL~\cite{Ade:2013zuv} and extract the relic abundance from Figure 2 of~\cite{Cirelli:2007xd}. The result extracted from Figure 2 of~\cite{Hisano:2006nn} has a thermal relic abundance for a mass closer to 3.1 TeV.} For lighter winos, due to their large tree-level annihilation rate to $W$ bosons, achieving the measured relic abundance requires a non-thermal history, e.g. the Moroi--Randall scenario with a late-decaying modulus~\cite{Moroi:1999zb}. It has been argued that moduli with mass of order ${\cal O} (100~{\rm TeV})$ are consistent with BBN and lead to the right relic abundance for light winos with mass about a few hundred GeV or lower~\cite{Moroi:1999zb,Kaplan:2006vm,Acharya:2009zt,Arcadi:2011ev}. Given that generically, moduli masses are set by the SUSY breaking scale and of order the gravitino mass~\cite{deCarlos:1993jw}, this non-thermal history fits nicely into the ``split SUSY'' scenarios with scalars and gravitino at about 100 TeV. Pure wino dark matter is challenging to directly detect due to its small cross section for scattering off nucleons, $\sigma_{p} \sim 10^{-47}$ cm$^2$~\cite{Hisano:2011cs, Hill:2011be}. On the other hand, it has a large self-annihilation cross section that current indirect dark matter searches are already sensitive to.

In general, it will be useful to study the implications of indirect as well as direct dark matter searches for some benchmark models of neutralino dark matter classified by dark matter composition and cosmological histories: 
 \begin{enumerate}
 \item{\bf{Single state neutralino dark matter}}
 \begin{enumerate}
 \item{Pure wino dark matter: thermal scenario with mass at about 2.8 TeV; non-thermal scenarios, such as the decaying moduli scenario for light winos with mass around a couple of hundred GeV or alternative scenarios like decaying gravitinos (we do not consider winos heavier than $\approx 3$ TeV, which in these well-motivated nonthermal cosmologies would have too large a relic abundance);}
 \item{Pure bino dark matter: thermal co-annihilation scenario with light sleptons; non-thermal scenarios with the late-decaying particles decaying dominantly to standard model particles (this cannot be realized in the moduli scenario or gravitino scenario); }
 \item{Pure higgsino dark matter: thermal scenario with mass at 1 TeV; non-thermal scenario for light higgsino similar to the ones discussed for pure wino case. (Strictly speaking, pure higgsinos are Dirac and have a large direct detection cross section, but we assume a very small mixing with a bino or wino, enough to split the two neutral Majorana higgsino mass eigenstates and forbid elastic scattering.)}
 \end{enumerate}
 \item {{\bf{Mixed state neutralino dark matter}}: bino/higgsino, wino/higgsino, or bino/wino/higgsino scenarios (bino and wino mix through higgsino). The thermal history is the well-tempered scenario~\cite{ArkaniHamed:2006mb} (for earlier references, see~\cite{Feng:2000gh, Giudice:2004tc, Pierce:2004mk}) while the non-thermal history could be one of or a mixture of those for the single states depending on the composition;}
 \item{{\bf{Multi-component dark matter}}: the neutralino is only one component of dark matter with either a thermal or non-thermal history while other candidates such as axions constitute the rest of the dark matter.}
\end{enumerate}
 In this paper, we will first study constraints of indirect searches for gamma rays on the relic abundance of wino dark matter, in the case that it is all of the dark matter or just one of the components of dark matter, and the associated astrophysical uncertainties on the constraints. Although the case of thermal wino dark matter is not our primary focus, we will assess its status; the implication of the HESS line search for the case of thermal winos was first noticed by the authors of Ref.~\cite{Cohennewpaper}, to which our work is complementary. The constraints we describe have profound implications for moduli masses and SUSY breaking scales in the non-thermal moduli decay scenario, which we will discuss in detail. We will leave the implications for non-thermal gravitino scenarios for future study. We will also comment on pure bino and higgsino scenarios and initiate a study of mixed states by showing that for the bino/higgsino scenario, future indirect searches could be complementary to direct searches. 

\section{Constraints on wino dark matter from indirect dark matter searches}
\label{sec:constraints}
We first present constraints on neutralino dark matter from indirect detection experiments which search for either excesses in the photon continuum spectrum or a line-like feature. The continuum photons arise mostly from fragmentation of hadronic final states in the tree-level processes ${\tilde \chi}^0 {\tilde \chi}^0 \to W^+ W^-$ (for winos and higgsinos) and ${\tilde \chi}^0 {\tilde \chi}^0\to ZZ$ (for higgsinos). Lines arise from the one-loop processes ${\tilde \chi}^0{\tilde \chi}^0 \to \gamma\gamma$ and ${\tilde \chi}^0{\tilde \chi}^0 \to Z\gamma$. We limit our attention to searches involving gamma rays, although searches in antiprotons set limits comparable to those in the photon continuum~\cite{Belanger:2012ta,Chun:2012yt}. Understanding the antiproton bounds requires more detailed astrophysics, related to how cosmic rays propagate through the galaxy. Because the gamma ray constraints are comparably strong, and more easily understood, we will exclusively focus on gamma rays. We will also neglect the possibility of sharp line-like spectral features arising from internal bremsstrahlung~\cite{Bringmann:2007nk,Birkedal:2005ep}, which require scalars tuned to be nearly degenerate in mass with the neutralino, and hence are absent in models with a split spectrum. Internal bremsstrahlung associated with ${\tilde \chi}^0 {\tilde \chi}^0 \to W^+ W^- \gamma$ gives a broader, subdominant feature that we will not take into account. We will initially discuss the bounds for standard cuspy (NFW and Einasto) dark matter halo profiles, before turning at the end of the section to a discussion of the possibility that the Milky Way halo has a large constant-density core of dark matter.

\subsection{Photon continuum constraints}
\label{sec:con}
\begin{figure}[!h]\begin{center}
\includegraphics[width=0.6\textwidth]{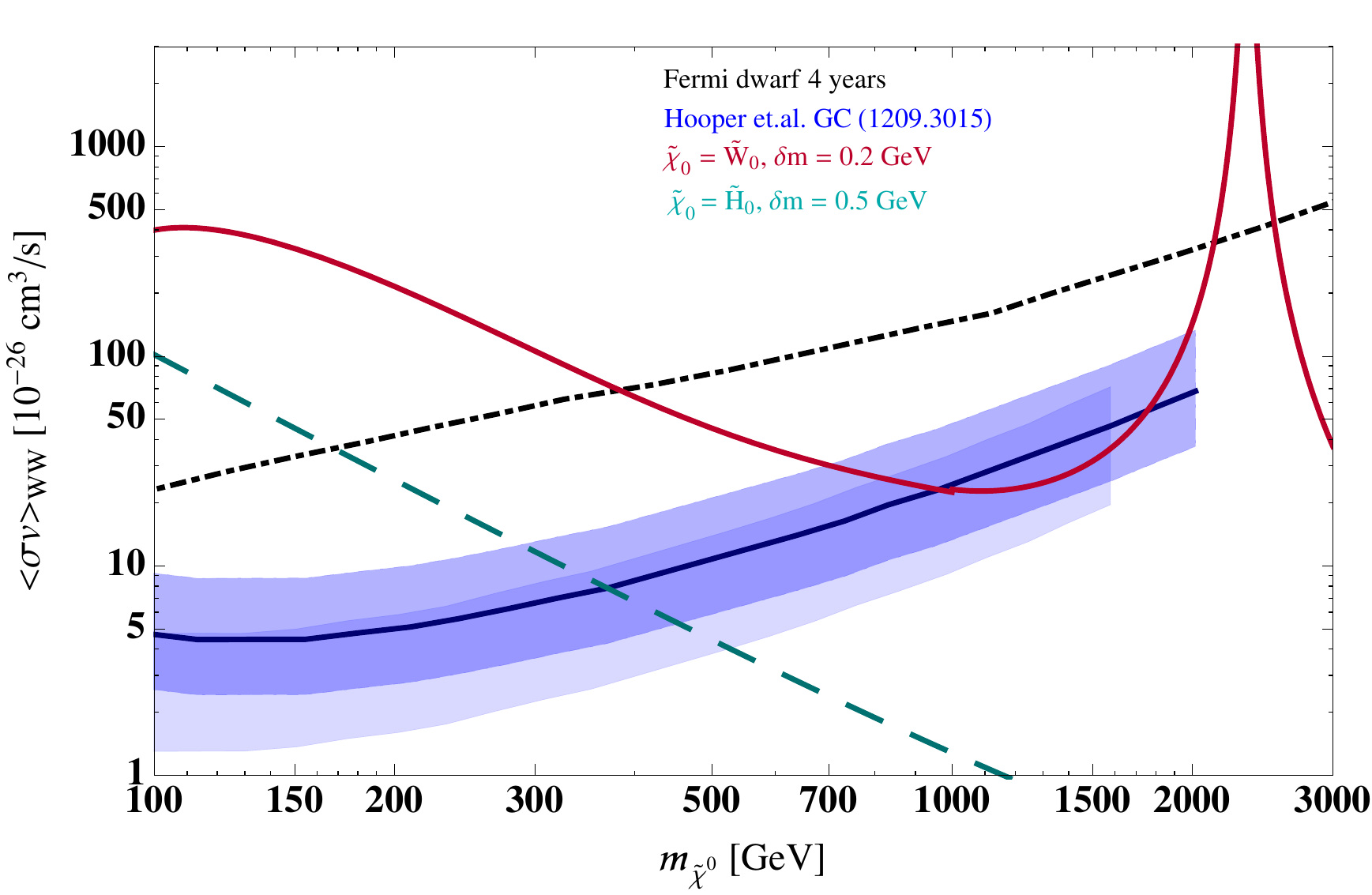}
\end{center}
\caption{Constraints on the cross section of annihilation into $WW (+ZZ)$ final state and wino/higgsino annihilation cross section as a function of neutralino mass. The black dot-dashed curve is the constraint from the continuum photon spectrum of Milky Way satellite galaxies~\cite{DrlicaWagner2012}; the dark blue curve is the constraint from the photon continuum in our galactic center assuming an NFW profile with $\rho(r_\odot) = 0.4$ GeV/cm$^3$ and $r_\odot = 8$ kpc~\cite{Hooper:2012sr}. The blue (lighter blue) bands are derived by varying $\rho(r_\odot)$ of NFW (Einasto) dark matter profiles as discussed in the text. The burgundy solid (cyan dashed) curve is the cross section of wino (higgsino) annihilation into $WW(+ZZ)$ final states.}
\label{fig:continuum}
\end{figure}

In this subsection, we summarize existing constraints from searches of the photon continuum in either the Milky Way's satellite dwarf galaxies~\cite{Ackermann:2011wa,DrlicaWagner2012} or the galactic center~\cite{Hooper:2012sr}. The results are presented in Fig.~\ref{fig:continuum}, in which we plotted the bounds on the cross section of continuum production from~\cite{DrlicaWagner2012, Hooper:2012sr}. The bound from satellite dwarf galaxies is based on Fermi 4 year data in the reprocessed Pass 7 Clean event class and is already marginalized over astrophysical uncertainties. The bound from the galactic center derived in~\cite{Hooper:2012sr} is based on Fermi 4 year data in the Pass 7 Ultraclean class, assuming some conservative choices of dark matter profiles. (A result from the HESS collaboration~\cite{Abramowski:2011hc}, also observing the galactic center, sets a similar constraint in the range above 500 GeV.) We rescale it by varying the overall constant factor in the dark matter profiles to illustrate its associated astrophysical uncertainties. The procedure is as follows. 

The flux of photons from dark matter annihilation is given by
\beq
\frac{d \Phi_\gamma}{d E_\gamma} &=& \frac{\langle \sigma v \rangle}{8 \pi m_{\chi}^2}\frac{dN_\gamma}{d E_\gamma} \int_{ROI} \frac{d J}{d \Omega} d \Omega, \\
J&=&\int_{ROI} d s d\Omega \; \rho(r)^2,
\eeq
where $\langle \sigma v \rangle$ and $m_\chi$ are the annihilation cross section and mass of dark matter. $\frac{dN_\gamma}{d E_\gamma}$ is the gamma-ray spectrum produced per annihilation. For the photon continuum, it depends on final-state radiation and the decays of hadrons (in particular, $\pi^0$'s) arising from fragmentation of the final state (often calculated via PYTHIA~\cite{Sjostrand:2000wi}) while for the monochromatic photon search, it is $N_\gamma \delta (E_\gamma - m_\chi)$. The $J$ factor is the integration over the region of interest (ROI) and line of sight of the square of the dark matter density profile. 

When setting the bounds using the data from our galactic center, we focus on two cuspy dark matter profiles supported by $N$-body simulations, namely the NFW (Navarro-Frenk-White) profile~\cite{Navarro:1995iw, Navarro:1996gj} and the Einasto profile~\cite{Navarro:2003ew, Springel:2008cc}:
\beq
\rho(r)&=&\frac{\rho_s}{(r/R_s)\left(1+(r/R_s)\right)^2} \quad {\rm NFW}, \\
\rho(r)&=&\rho_s \exp\left[\frac{-2}{\alpha}\left(\left(\frac{r}{R_s}\right)^\alpha-1\right)\right]\quad {\rm Einasto},
\eeq
where the scale radius of the halo $R_s = 20$ kpc and for the Einasto profile, we use $\alpha = 0.17$. The characteristic density $\rho_s$ is determined by the local dark matter density at the sun, $\rho(r_\odot)$. The distance between the sun and the galactic center is taken to be $r_\odot = 8$ kpc throughout this paper.\footnote{A popular choice of $r_\odot$ in setting the bounds is 8.5 kpc, which corresponds to larger $\rho_s$ for fixed $\rho(r_\odot)$ and thus stronger bounds.}

Now we illustrate the uncertainties in calculating the $J$ factor by varying $\rho(r_\odot)$ and thus $\rho_s$ in dark matter profiles. A recent study of microlensing and dynamical observations of our galaxy mapped out 2$\sigma$ boundaries of $\rho(r_\odot)$ for NFW and Einasto profiles~\cite{Iocco:2011jz}. For the NFW profile, $\rho(r_\odot) = 0.29 - 0.54~{\rm GeV/cm}^3$ and for the Einasto profile, $\rho(r_\odot) = 0.25 - 0.48~{\rm GeV/cm}^3$ at the 2$\sigma$ level fixing $r_\odot = 8$ kpc. 
We rescale the bounds in~\cite{Hooper:2012sr}, which used exclusively the lower (conservative) end of the 2$\sigma$ range of Ref.~\cite{Iocco:2011jz}, and plot the band of bounds in Fig.~\ref{fig:continuum} by varying $\rho(r_\odot)$ in the 2$\sigma$ ranges listed above. As the Einasto profile has a steeper inner slope than NFW, it leads to a bigger $J$ factor and thus a stronger bound for searches concentrating near the galactic center, all else being equal. This is demonstrated in Fig.~\ref{fig:continuum}, in which the lighter shaded band of bounds from Einasto profiles is lower than the darker shaded band of bounds from NFW profiles. 
In Fig.~\ref{fig:continuum}, we also plot the bound assuming an NFW profile with $\rho(r_\odot) = 0.4$ GeV/cm$^3$, which is a common value used in setting bounds, as a darker reference curve. In Sec.~\ref{sec:Jfactor}, we will discuss dark matter profiles with softened cusps such as cored profiles. 

We also present the production cross section as a function of neutralino dark matter mass in Fig.~\ref{fig:continuum}. In calculating the wino annihilation cross section, we take into account the Sommerfeld enhancement and one-loop corrections using the fitting functions in~\cite{Hryczuk:2011vi}. For the higgsino annihilation cross section, we use fitting functions in~\cite{Hisano:2004ds}, which only take into account Sommerfeld enhancement. For the plot, the splitting between the charged and neutral winos is set to be 0.2 GeV and the higgsino mass splitting is 0.5 GeV. These are reasonable approximations to the expected splittings, which have little effect on these tree-level rates. We review the physics of these mass splittings in Appendix~\ref{app:masssplitting}. 

From Fig.~\ref{fig:continuum}, one could see that conservatively, the Fermi dwarf galaxy data rules out pure wino dark matter up to around 385 GeV and pure higgsino dark matter up to around 160 GeV. The dwarf galaxy data also rules out wino dark matter with mass around 2.4 TeV, where the first resonance enhancement peak lies. The galactic center photon continuum data rules out wino dark matter up to around 700 GeV and higgsino dark matter up to 300 GeV for either NFW or Einasto profiles. 

\subsection{Photon line constraints}
Both Fermi and HESS searches for line-like features in the photon spectrum are already sensitive to the cross section of wino dark matter annihilating into two photons or a photon and a $Z$ boson~\cite{Fermi-LAT:2013uma,Abramowski:2013ax}. The difference is that currently the Fermi search is only sensitive to photons with energy below 300 GeV, while HESS is sensitive to photons in a higher energy range above 500 GeV. In this subsection, we will derive bounds on neutralino dark matter annihilation from photon line searches. 

\subsubsection{Neutralino annihilations into two photons}
\begin{figure}[!h]\begin{center}
\includegraphics[width=0.5\textwidth]{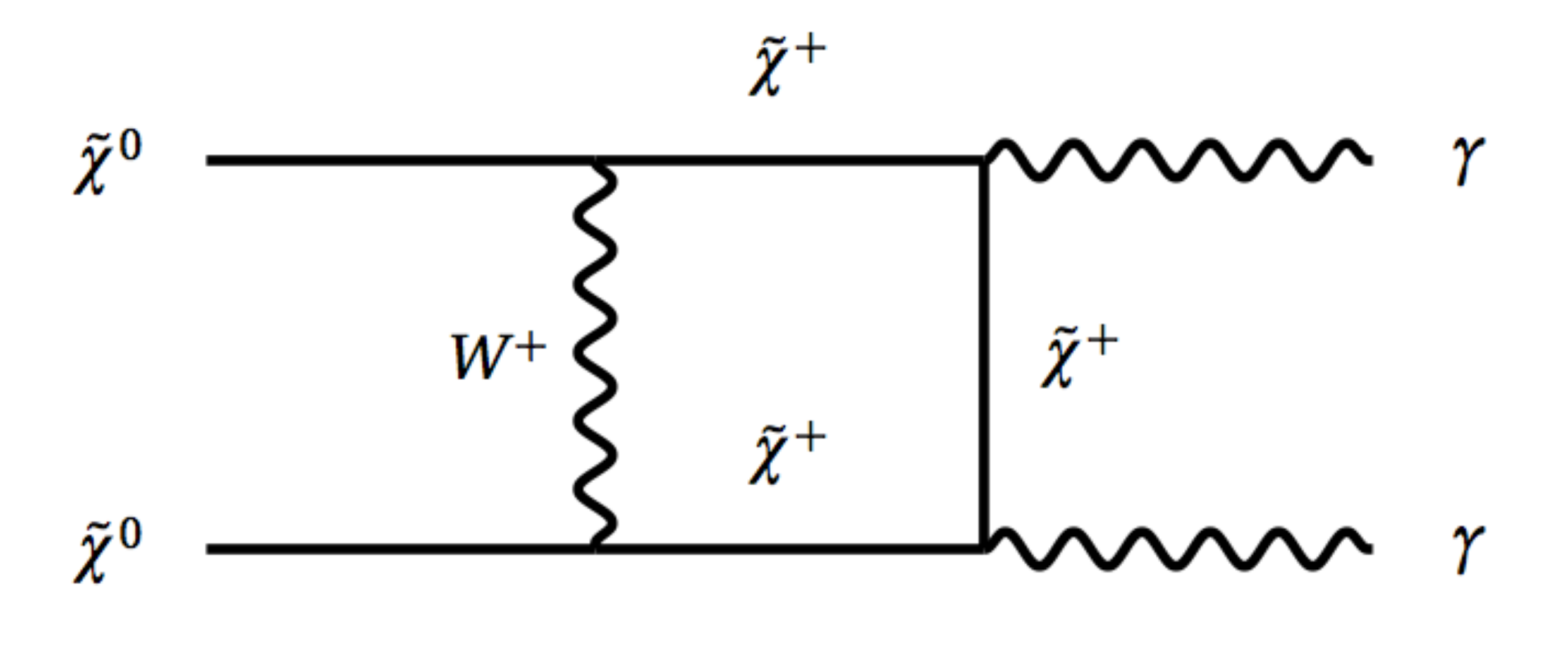}
\end{center}
\caption{Dominant diagram in the wino or higgsino annihilation into photons at the one-loop level, in the limit when the neutralino is heavy.}
\label{fig:diagram}
\end{figure}

Analytic results of the full one-loop calculation of neutralino annihilation into two photons or photon+$Z$ have been derived in~\cite{Bergstrom:1997fh, Ullio:1997ke, Bern:1997ng, Boudjema:2005hb}. The Sommerfeld enhancement for pure wino or pure higgsino have been calculated in~\cite{Hisano:2004ds,Hryczuk:2011vi}. The two calculations are different and there are some limitations of both calculations, which we will discuss in Appendix~\ref{app:winomatching}. To understand the behavior of the cross sections, we first inspect the limit when the neutralino is heavy and the lightest superpartner (LSP) and its corresponding charged state are nearly degenerate in masses. We will neglect Sommerfeld enhancement for the moment. In this limit, only one type of box diagram dominates, as shown in Fig.~\ref{fig:diagram}. Other contributions to the rate are suppressed by $1/m_\chi^2$.
The analytic formula of the cross sections in this limit are given by  
\beq
\langle \sigma v \rangle_{\tilde{\chi}^0\tilde{\chi}^0\to \gamma\gamma}&\approx&\frac{4\alpha^4 \pi}{m_W^2 \sin^4\theta_W} \approx1.6 \times 10^{-27}\,{\rm cm}^3/s  \quad(\tilde{\chi}^0=\tilde{W}^0) , \nonumber\\
&\approx& \frac{\alpha^4 \pi}{4m_W^2 \sin^4\theta_W} \approx 10^{-28} \,{\rm cm}^3/s\quad  (\tilde{\chi}^0=\tilde{H}^0) , \\
\langle \sigma v \rangle_{\tilde{\chi}^0\tilde{\chi}^0\to Z\gamma}&\approx&\frac{8\alpha^4 \pi \cos^2\theta_W}{m_W^2 \sin^6\theta_W} \approx1.1 \times 10^{-26}\,{\rm cm}^3  /s\quad(\tilde{\chi}^0=\tilde{W}^0) , \nonumber\\
&\approx&\frac{\alpha^4 \pi \left(\sin^2\theta_W-0.5\right)^2}{2m_W^2 \sin^6\theta_W\cos^2\theta_W} \approx8.0\times 10^{-29}\,{\rm cm}^3/s  \quad(\tilde{\chi}^0=\tilde{H}^0) .
\label{eq:noenhance}
\eeq
We see that for heavy neutralino, without Sommerfeld enhancement, its annihilation cross section is approximately a constant, independent of its mass at the leading order. (Taking into account the small but finite mass splitting leads to a gradual decline in this cross section at high masses.)

For pure winos, the $Z \gamma$ annihilation cross section is about one order of magnitude larger than $\gamma\gamma$ annihilation, whereas for pure higgsinos they are comparable. The differences in wino and higgsino production cross sections originate from their couplings to $Z$ and $\gamma$. For a $\gamma\gamma$ final state, there is an additional Bose factor of $1/2$ compared to $Z\gamma$.

\begin{figure}[!h]\begin{center}
\includegraphics[width=0.8\textwidth]{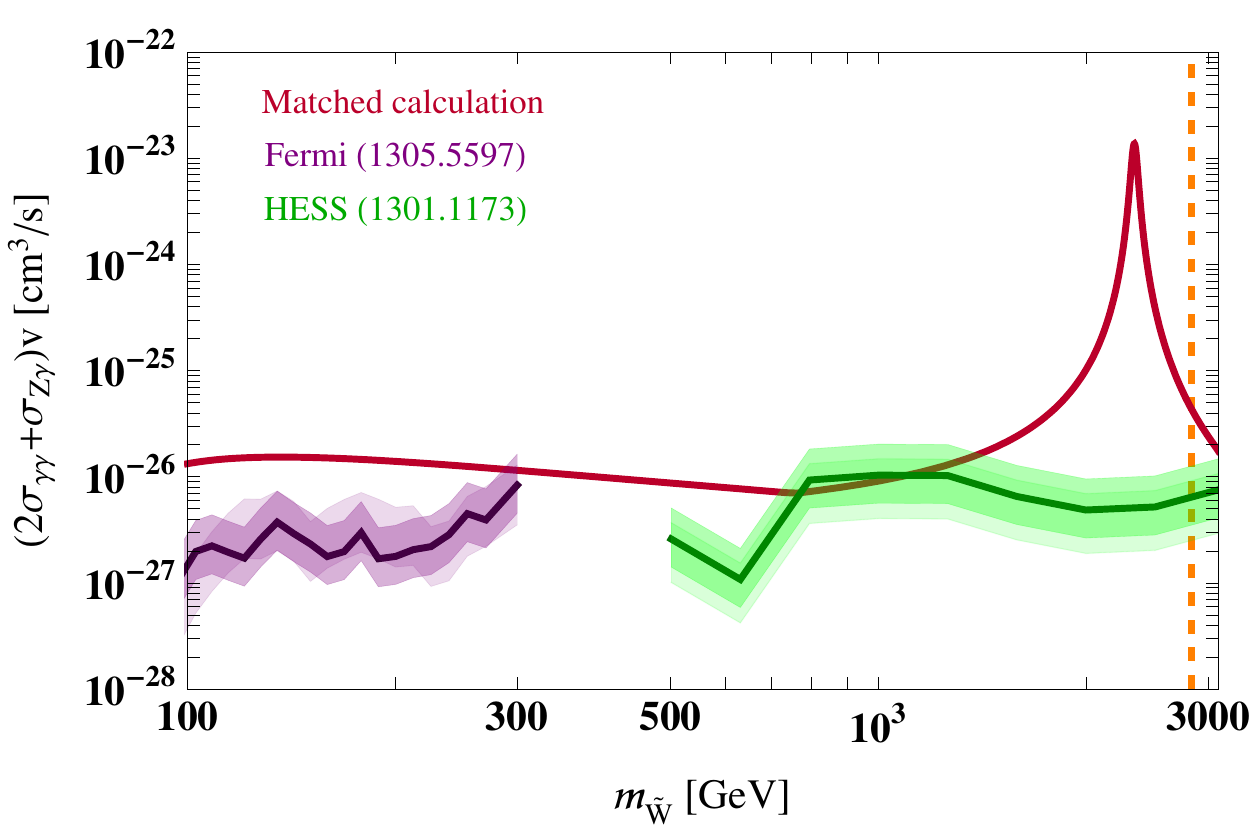}
\end{center}
\caption{Constraints on the cross section of wino annihilation into photon(s). The burgundy solid curve is the wino annihilation cross section by matching one-loop calculation~\cite{Bergstrom:1997fh, Ullio:1997ke, Bern:1997ng, Boudjema:2005hb} and the Sommerfeld enhancement calculation~\cite{Hryczuk:2011vi}. Details can be found in Appendix~\ref{app:winomatching}. The purple curve is the constraint from the Fermi line search~\cite{Fermi-LAT:2013uma} assuming an NFW profile with $\rho(r_\odot) = 0.4$ GeV/cm$^3$ and $r_\odot = 8$ kpc. The purple (lighter purple) bands are derived by varying $\rho(r_\odot)$ of NFW (Einasto) dark matter profiles as discussed in the text. The green curve is the constraint from the HESS line search~\cite{Abramowski:2013ax} assuming an NFW profile with $\rho(r_\odot) = 0.4$ GeV/cm$^3$ and $r_\odot = 8$ kpc. The green (lighter green) bands are derived by varying $\rho(r_\odot)$ of NFW (Einasto) dark matter profiles as discussed in the text. The vertical dashed orange line marks the wino with thermal relic abundance $\Omega_{\rm thermal} h^2 =0.12$.}
\label{fig:line}
\end{figure}

In Fig.~\ref{fig:line}, we plotted the total cross section of wino annihilation into photons weighted by the number of photons in the final state, $2\langle \sigma v\rangle_{\gamma\gamma} +\langle \sigma v\rangle_{Z\gamma}$, as a function of the wino mass. The cross section is a result of matching between the one-loop analytic calculation, which is more reliable for light winos, and the calculation including Sommerfeld enhancement, which kicks in around a TeV wino mass. The details of the calculations and matching between different calculations can be found in Appendix~\ref{app:winomatching}. We have not plotted the higgsino annihilation rate, which is too small for current experiments to exclude.

\subsubsection{Constraints from Fermi and HESS line searches}
Both the Fermi and HESS collaborations have reported dark matter constraints from photon line searches in the galactic center~\cite{Fermi-LAT:2013uma, Abramowski:2013ax}. The constraints rule out a cross section $\langle \sigma v \rangle \sim 10^{-27} - 10^{-26}$ cm$^3$/s depending on the dark matter mass. The quantitative bounds are presented in Fig.~\ref{fig:line}. The Fermi line search defined four regions of interest for annihilating dark matter, with each region optimized for a particular dark matter halo profile. The HESS line search has one search region of interest contained within a 1$^\circ$ circle near the galactic center, and hence is weakened more for less concentrated halo profiles. Both Fermi and HESS analyses assumed $r_\odot = 8.5$ kpc and $\rho(r_\odot) = 0.4$ GeV/cm$^3$. 

To have a unified normalization of dark matter profiles and estimate the astrophysical uncertainties, we followed the same strategy we used in setting the bounds from continuum photons in the galactic center as discussed in Sec.~\ref{sec:con}. Again we only focused on cuspy profiles, i.e., NFW and Einasto profiles, in this section. In Fig.~\ref{fig:line}, we rescale the bounds in~\cite{Fermi-LAT:2013uma,  Abramowski:2013ax} and plot the bounds assuming the NFW profile with $\rho(r_\odot) = 0.4$ GeV/cm$^3$ and $r_\odot = 8$ kpc as reference curves. We also plot the bands of bounds in Fig.~\ref{fig:line} by varying $\rho(r_\odot)$ in the 2$\sigma$ range from~\cite{Iocco:2011jz}. Notice that for the Fermi line constraints, the NFW band and Einasto band have different shapes because the Fermi line analysis used different search regions for NFW and Einasto profiles. In Sec.~\ref{sec:Jfactor}, we will discuss dark matter profiles with softened cusps such as cored profiles. 

In setting the bounds, we neglected the energy differences of photons in $\gamma\gamma$ and $\gamma Z$ final states for $m_{\tilde{\chi}^0} \geq 200$ GeV, assuming the two final states contribute to a single line-like feature in the fit. The energy of the photon in the $\gamma Z$ final state is larger than that of the photons in $\gamma\gamma$ by an amount 
\beq
\delta m = \frac{m_Z^2}{4 m_{\tilde{\chi}^0}} \approx 10\, {\rm GeV} \left(\frac{200 {\rm GeV}}{m_{\tilde{\chi}^0}}\right)^2.
\eeq
Given the current energy resolutions of both experiments $\gsim 10$ GeV, this is a reasonable approximation for $m_{\tilde{\chi}^0} \geq 200$ GeV~\cite{Ackermann:2012kna, Abramowski:2013ax}. For $100 \,{\rm GeV}\leq m_{\tilde{\chi}^0} < 200 \,{\rm GeV}$, we consider only the contribution of the process ending in $\gamma Z$ to the photon line flux because it is about $2.5 - 2.8$ times that of the process leading to $\gamma\gamma$. 

From Fig.~\ref{fig:line}, we can see that if dark matter is purely wino, the constraint from line searches rules out winos in the range $(100-300)$ GeV and (500 GeV$-3$ TeV), with (700 GeV$-1.4$\; TeV) less constrained or unconstrained depending on the astrophysical parameters. {\bf Combined with constraints from continuum photons from galactic center, pure wino dark matter in the whole range from 100 GeV to 3 TeV (with the possible exception of a range between 700 GeV and 1.4 \;TeV) is ruled out for both NFW and Einasto profiles, allowing astrophysical parameters to vary in the 2$\sigma$ range in~\cite{Iocco:2011jz}.} 
 
\begin{figure}[!h]\begin{center}
\includegraphics[width=0.8\textwidth]{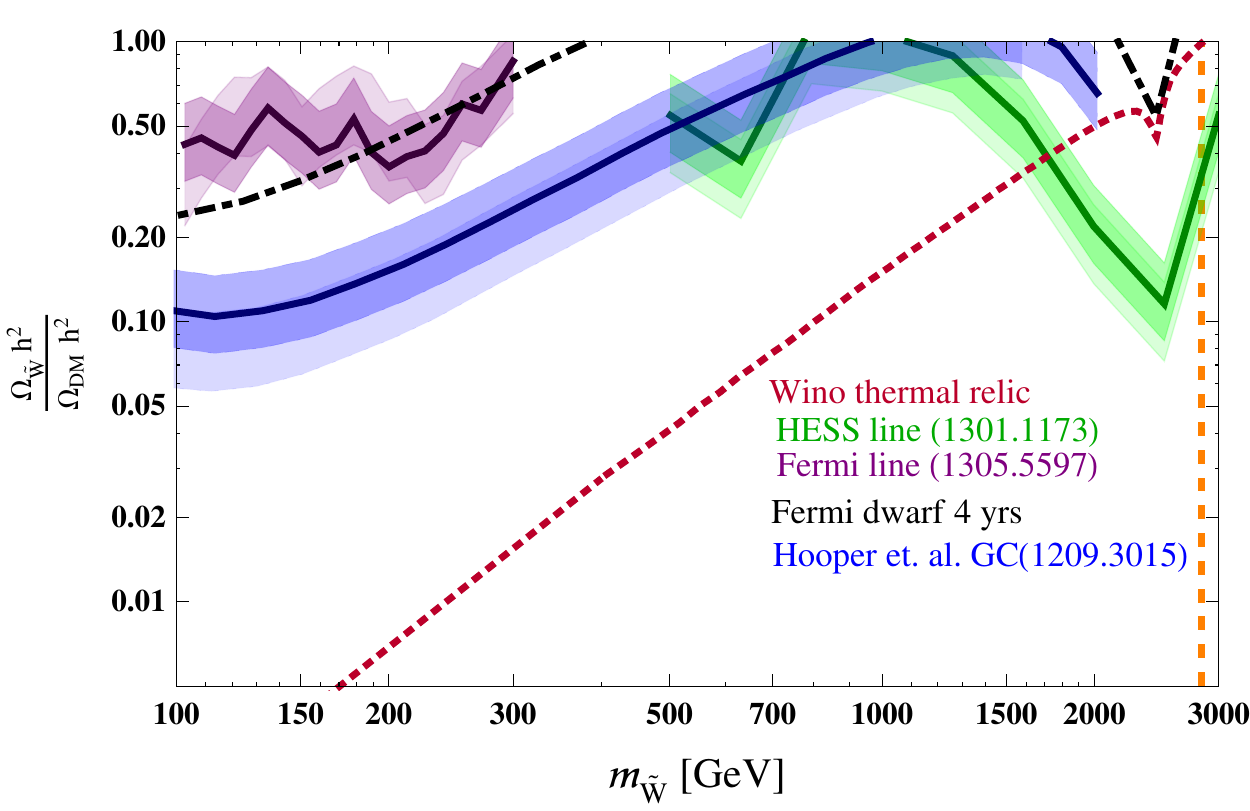}
\end{center}
\caption{Constraints on the relic abundance of wino dark matter (i.e., a wino component in a scenario with multiple dark matter particles). The burgundy dashed curve is the thermal relic abundance of winos calculated in~\cite{Hisano:2006nn, Cirelli:2007xd}. The other curves are constraints from different indirect detection searches. Black dot-dashed: Fermi dwarf galaxy; purple line and bands: Fermi line search assuming NFW profile with $\rho(r_\odot) = 0.4$ GeV/cm$^3$ with $r_\odot = 8$ kpc (purple solid line), NFW profile with varying $\rho(r_\odot)$ (purple band), Einasto profile with varying $\rho(r_\odot)$ (lighter purple band); green line and bands: HESS line search assuming NFW profile with $\rho(r_\odot) = 0.4$ GeV/cm$^3$ with $r_\odot = 8$ kpc (green solid line), NFW profile with varying $\rho(r_\odot)$ (green band), Einasto profile with varying $\rho(r_\odot)$ (lighter green band); blue line and bands: Fermi galactic center continuum search analyzed in~\cite{Hooper:2012sr} assuming NFW profile with $\rho(r_\odot) = 0.4$ GeV/cm$^3$ with $r_\odot = 8$ kpc (blue solid line), NFW profile with varying $\rho(r_\odot)$ (blue band), Einasto profile with varying $\rho(r_\odot)$ (lighter blue band). The vertical dashed orange line marks the wino with thermal relic abundance $\Omega_{\rm thermal} h^2 =0.12$.}
\label{fig:reliclimit}
\end{figure}%

In Fig.~\ref{fig:reliclimit}, we present constraints from various indirect searches using photons on the relic abundance of a wino dark matter component. In the plot, we also plotted the wino thermal relic abundance calculated in~\cite{Hisano:2006nn, Cirelli:2007xd}. From Fig.~\ref{fig:reliclimit}, the wino dark matter scenario with a thermal relic equal to the observed dark matter relic, which we took to be $\Omega h^2 = 0.12$~\cite{Ade:2013zuv}, is ruled out for NFW or Einasto profiles. Below 1.5 TeV, the bound on the allowed relic abundance of winos is above the thermal relic abundance, and thus a non-thermal contribution to the wino relic abundance is still allowed but is bounded to be less than all of the dark matter. 

\subsection{Core vs cusp dark matter profiles}
\label{sec:Jfactor}

Numerical simulations of galaxy formation including only dark matter robustly find cuspy dark matter distributions like the NFW and Einasto profiles we have discussed so far. Of course, the inner region of the Milky Way galaxy is not solely composed of dark matter; sufficiently near the center, the galaxy is dominated by baryons. The effect of baryons on the shapes of dark matter halos is still uncertain. Even the {\em sign} of the effect is in dispute. Adiabatic contraction tends to make the dark matter profiles steeper in the galactic center, as argued on theoretical grounds~\cite{Blumenthal:1985qy} and observed in simulations (e.g.~\cite{Gnedin:2011uj}). If this is the dominant effect, it will tend to increase indirect detection signals from the galactic center, and by ignoring it we are being conservative. However, baryons could also lead to dark matter distributions without cusps, a possibility that has drawn a great deal of attention in the context of dwarf satellite galaxies, which appear to have cored halos. Feedback from supernovae, for instance, has been suggested as a possible culprit in the destruction of cusps. Recent high-quality numerical simulations producing realistic spiral galaxies have found that cusps survive even repeated baryonic outflows~\cite{Marinacci:2013mha}. Perhaps the most dangerous effect for the interpretation of indirect detection limits is a resonant bar/halo interaction, which may lead to formation of a core of kiloparsec size in the Milky Way~\cite{Weinberg:2001gm}. Recent work has argued that the Eris simulation shows evidence for a 1 kpc core in the Milky Way~\cite{Kuhlen:2012qw}, in contrast to earlier work arguing that core formation was an artifact of simulations with too large a timestep~\cite{Dubinski:2008yi}. On the other hand, one of the simulated galaxies in Ref.~\cite{Marinacci:2013mha} has a prominent bar and does not have a core. In short, the $N$-body simulation community does not appear to have converged on an answer for the expected shape of the Milky Way's inner halo. Observations also offer little help; a recent fit claimed a mild preference for a large core~\cite{Nesti:2013uwa}, but was also compatible with an NFW-like distribution. It is also worth keeping in mind that even if observations decisively favored a cored profile, this would not necessarily be good news for proponents of wino dark matter. We would still face the question of whether cold dark matter with baryonic feedback could produce such a core; if not, the core might well point to self-interacting dark matter or other new dynamics incompatible with winos. Indeed, in the case of dwarf galaxies, the ``core/cusp problem'' is often cited as a motivation for moving beyond the paradigm of cold, collisionless dark matter.

\begin{figure}[!h]\begin{center}
\includegraphics[width=0.5\textwidth]{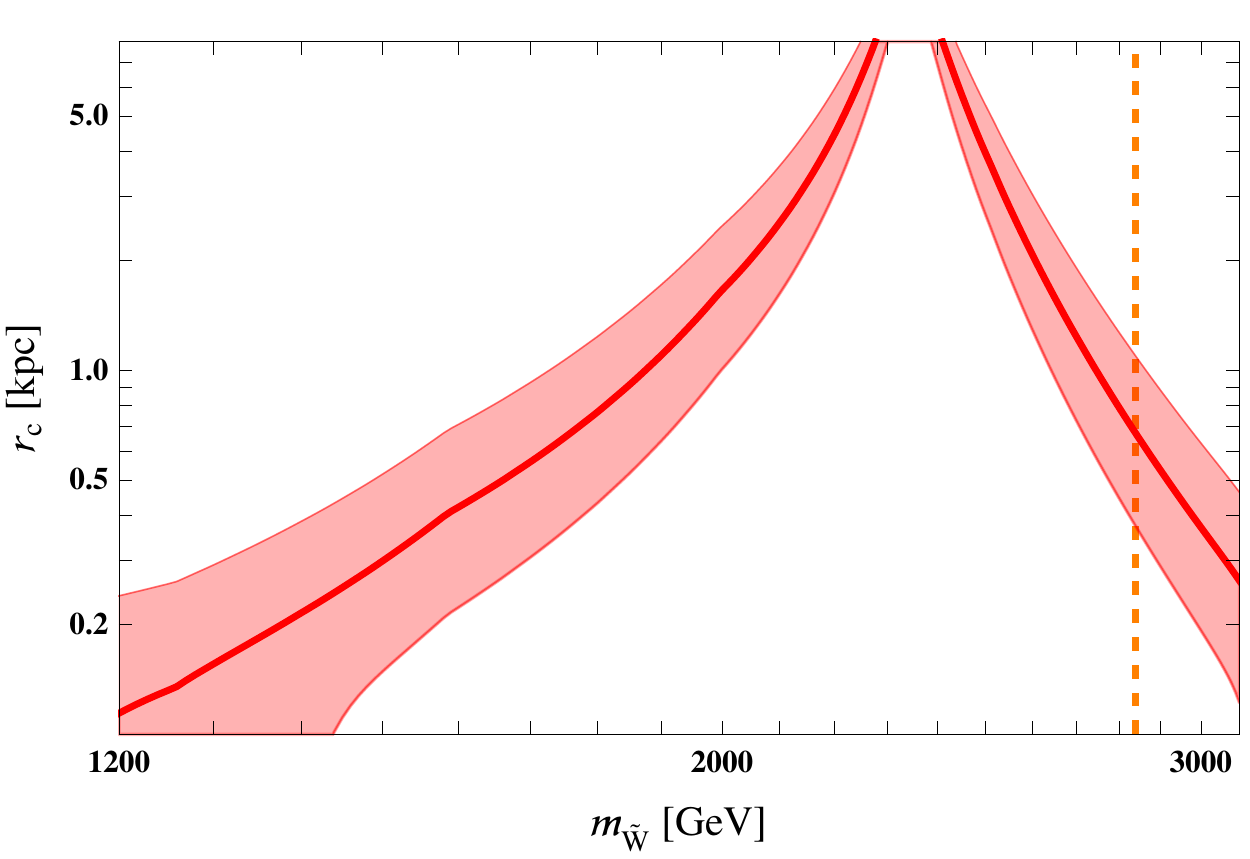}\includegraphics[width=0.5\textwidth]{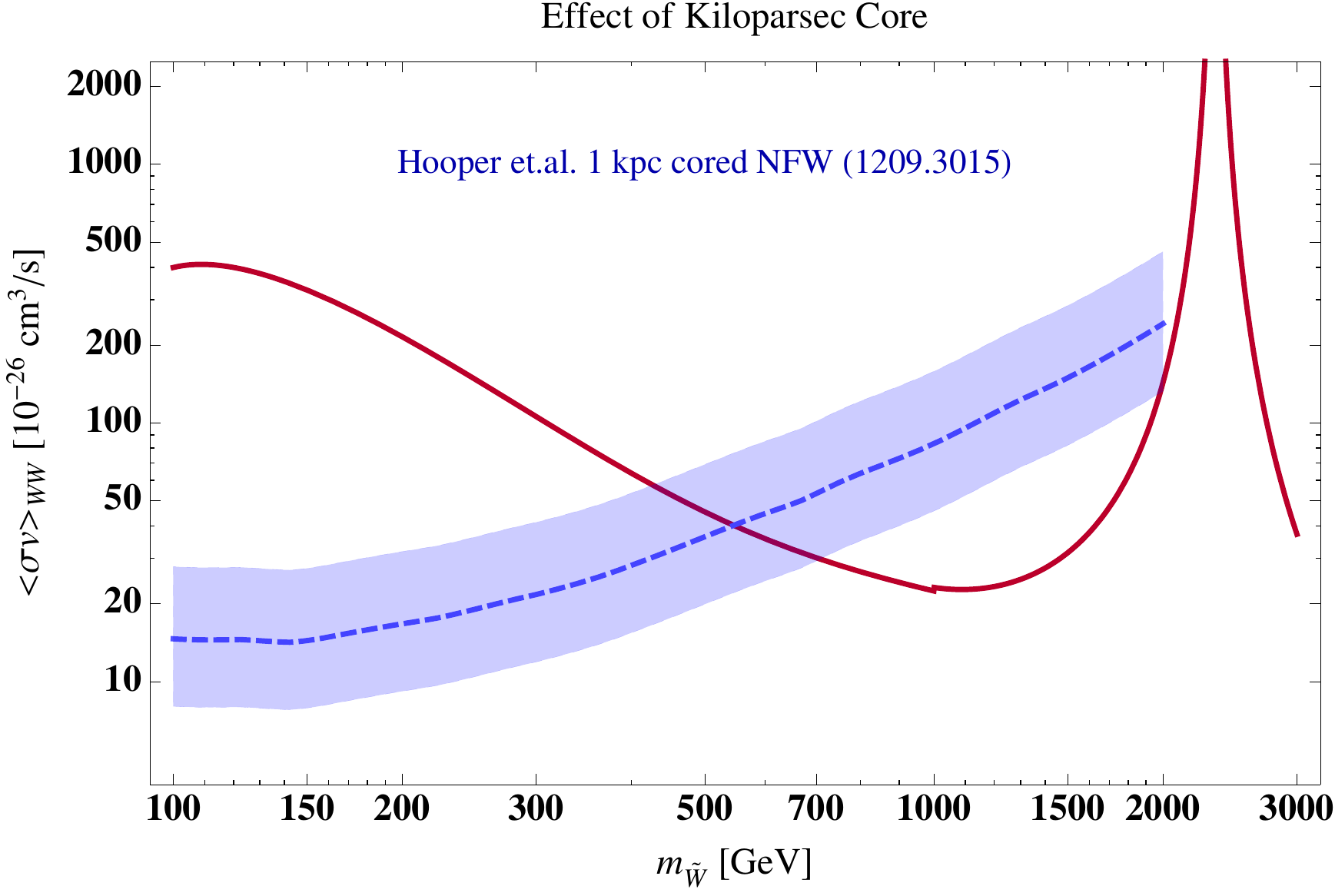}
\end{center}
\caption{Left: minimal radius of the inner constant density core which will remove the HESS limits as a function of wino mass. The band is obtained by varying $\rho(r_\odot)$ in the range $0.29 - 0.54$ GeV/cm$^3$. The solid red reference curve corresponds to $\rho(r_\odot) = 0.4$ GeV/cm$^3$.
The vertical dashed orange line marks the wino with thermal relic abundance $\Omega_{\rm thermal} h^2 =0.12$. Right: the bound from Ref.~\cite{Hooper:2012sr} in the case of an NFW profile with 1 kpc constant density core (blue band), compared to the expected wino cross section (burgundy curve).}
\label{fig:core}
\end{figure}%

Ideally, given this state of uncertainty, we would specify a prior on the space of halo shapes and marginalize over them to obtain a limit. Unfortunately, at this point, it is not clear that any uncontroversial prior can be chosen. As a result, we will limit ourselves to assessing how a cored distribution might alter the bounds. To study how a cored profile could affect the indirect search constraints, we consider an NFW-like profile with a flat, constant density core of radius $r_c$:
\beq
\rho_c(r)=  \begin{cases} \frac{\rho_s}{(r/R_s)\left(1+(r/R_s)\right)^2}  & r>r_c \nonumber \\
\frac{\rho_s}{(r_c/R_s)\left(1+(r_c/R_s)\right)^2} & r\leq r_c \end{cases}
\eeq
The cored profile leads to strong density suppression in regions of interest with radius smaller than $r_c$ and could change the constraints dramatically. As an example, we examine the core radius needed to relax the HESS line bound in the wino mass range $1.2 - 3$ TeV shown in Fig.~\ref{fig:reliclimit}. We choose $R_s = 20$ kpc, $\rho_c(r_\odot) = 0.4$ GeV/cm$^3$ and treat $r_c$ as a free parameter. In the left-hand side of Fig.~\ref{fig:core}, we plot the minimal $r_c$ at which the HESS limit allows winos to constitute all of the dark matter, as a function of wino mass. As expected, the HESS constraint can disappear only when the core radius is (much) larger than the HESS search region radius, which is around 0.1 kpc. More specifically, to allow thermal wino dark matter, the core radius has to be bigger than 0.4 kpc for $\rho (r_\odot) = 0.29$ GeV/cm$^3$ or 1 kpc for $\rho (r_\odot) = 0.54$ GeV/cm$^3$. We also present bound from continuum photons from the galactic center assuming a cored NFW profile with core radius $r_c = 1$ kpc in the right panel of Fig.~\ref{fig:core}, as extracted from~\cite{Hooper:2012sr}. Compared to bounds with NFW and Einasto profiles, the bound is relaxed to $(400-700)$ GeV for $\rho(r_\odot) = 0.29 - 0.54$ GeV/cm$^3$. Because it is uncertain whether the Milky Way has a cored profile, we leave the reader to decide between the following two possibilities: pure wino dark matter is ruled out in most of the mass range (assuming NFW or Einasto profiles from $N$-body simulations) or the Milky Way actually has a cored dark matter profile with a large core radius. If the second statement is true, this will pose potentially another cusp/core problem in the Milky Way in addition to the cusp/core problem in the dwarf galaxies. 

The Fermi line search and the continuum photon analysis of Ref.~\cite{Hooper:2012sr} look over a wide swath of the sky, and hence are still able to set strong (albeit weakened) limits in the case of a cored profile. The HESS bound at high masses is the most easily evaded, because it is based on a very small region of the sky near the galactic center. In part this is a consequence of using a Cherenkov telescope on the ground rather than a satellite like Fermi. However, given the importance for particle physics of understanding whether the thermal wino dark matter scenario is viable, we suggest that it might be worthwhile for HESS to reduce the experimental uncertainties by defining a secondary region of interest for photon line searches slightly away from the galactic center, which could constrain less cuspy profiles. We leave a detailed consideration of the best feasible strategy for future work.

\subsection{Complementary collider probe}
As an aside, we would like to emphasize that collider searches for disappearing tracks provide a complementary probe of wino dark matter~\cite{Feng:1999fu} in addition to indirect searches. The current result from ATLAS using 4.7 fb$^{-1}$ at 7 TeV only rules out winos with mass below 105 GeV~\cite{ATLAS:2012jp}. Further results will provide crucial information to solidify or weaken the constraints from indirect searches: null results of disappearing tracks could support the constraints of indirect searches derived using cuspy profiles, while a signal in this channel at colliders may suggest that indirect constraints should be alleviated by adopting a dark matter profile with a large core at the galactic center. On the other hand, a collider signal could also arise if winos are unstable on lifetimes long relative to the size of a detector but short on cosmological scales.
 
\section{Implications for non-thermal cosmology history}
\label{sec:moduli}
In this section, we will mainly focus on the implications of indirect search bounds for the scenario in which the relic abundance of winos is non-thermal and the winos are produced from decays of moduli fields. This is a representative scenario where late-decaying particles have a considerable branching fraction into dark matter, which annihilates sufficiently quickly to stay in chemical equilibrium until the decays stop. It is also well-motivated from a top-down viewpoint, as moduli appear ubiquitously in string constructions. In Appendix~\ref{app:non-thermal2}, we will discuss another possible type of non-thermal scenario where the late-decaying particles have a negligible branching fraction into wino dark matter. 

Non-thermal production of neutralinos from the decays of gravitinos is another interesting possibility which we will leave for future work.

\subsection{Basics of moduli decays}
 Moduli, scalar fields with Planck-suppressed couplings, are ubiquitous in string theory constructions, which have no tunable parameters. Coupling constants, like the $1/g^2$ factor in front of gauge field kinetic terms, always arise as VEVs of holomorphic gauge kinetic functions that depend on the moduli. Moduli fields begin to oscillate coherently around the minimum of their potential at Hubble scales of order their mass, leading to a matter-dominated phase of the universe that ends when moduli decay, reheating the universe to a radiation-dominated phase with temperature
  \beq
 T_{\rm RH} \equiv \left(\frac{90}{\pi^2 g_*(T_{\rm RH})}\right)^{1/4} \sqrt{\Gamma_{\phi} M_{\rm Pl}},
 \label{eq:Treheatdef}
 \eeq
where $M_{\rm Pl}$ is the reduced Planck mass $\approx 2.4 \times 10^{18}~{\rm GeV}$ and $\Gamma_\phi$ is the width of the decaying modulus. If $T_{\rm RH}$ is too low, e.g., below 5 MeV, this is in conflict with BBN, leading to a severe cosmological problem~\cite{Coughlan:1983ci,deCarlos:1993jw,Randall:1994fr}. One solution to this problem is that moduli fields have a large enough decay width $\Gamma_{\rm \phi}$ that their decays automatically reheat the universe to a high enough temperature for BBN to proceed normally, typically requiring $m_\phi \gsim 10$ TeV~\cite{Moroi:1994rs}. In this case, the decays of the moduli can produce dark matter, a scenario that fits best with light wino dark matter, given its large annihilation cross section~\cite{Moroi:1999zb,Kaplan:2006vm,Gelmini:2006pw,Acharya:2008bk,Acharya:2009zt,Kane:2011ih}. 

Note that the important physical parameter, which enters for instance in the Boltzmann equations, is $\Gamma_\phi$. Because the moduli do not all decay instantaneously, the decay happens over a range of temperatures around $T_{\rm RH}$, and conventions for characterizing this temperature differ; for example, Ref.~\cite{Moroi:2013sla} defines a ``decay temperature'' which is smaller by a factor of $9^{1/4}$. Our convention for defining $T_{\rm RH}$ is in agreement with the original paper of Moroi and Randall~\cite{Moroi:1999zb}, and we will plot the bound on $T_{\rm RH}$ in order to facilitate comparison with existing results. Yet it is more meaningful to discuss constraints on unambiguous physical quantities such as the moduli width and mass. The masses of moduli fields are typically quite closely connected to the SUSY breaking scale $m_{3/2}$. This is true even when the moduli are stabilized supersymmetrically, because tuning to cancel the cosmological constant relates the size of various terms in the superpotential, indicating that $m_\phi \gg m_{3/2}$ is possible only with additional fine-tuning~\cite{Fan:2011ua}. As a result, we expect that bounds on the reheat temperature are quite closely linked to a determination of the overall scale of supersymmetry breaking, in the absence of additional fine-tuning. To relate our bound on the reheat temperature to the moduli mass scale, we require an estimate for decay width. Moduli couple through Planck-suppressed operators, so
 \beq
 \Gamma_\phi = \frac{c}{4\pi} \frac{m_\phi^3}{M_{\rm Pl}^2}.
 \label{eq:moduliwidth}
 \eeq
Computing the coefficient $c$ requires a detailed model, but we can offer some general comments. In the case of a coupling of a modulus $\phi$ to an SU($N$) field strength, $\int d^2\theta \frac{\lambda}{M_{\rm Pl}} \phi \WW_\alpha \WW^\alpha + {\rm h.c.}$, the modulus has a decay width to gluons~\cite{Moroi:1999zb}
\beq
\Gamma(\phi \to gg) = \frac{\left(N^2-1\right)\left|\lambda\right|^2}{8\pi} \frac{m_\phi^3}{M_{\rm Pl}^2}.
\eeq
If $\phi$ is stabilized supersymmetrically, $F_\phi \propto m_\phi \phi$ and the decay rate to gluinos is precisely equal to that to gluons~\cite{Kaplan:2006vm,Endo:2006zj}. \textbf{There is no chirality suppression of moduli decaying to two gauginos.} The value of $\lambda$ will be order one in some cases (e.g. cases where $K \propto M_{\rm Pl}^2\log(T+T^\dagger)$, the gauge kinetic function is $\propto T$, and $\phi$ is the canonically-normalized fluctuation around the VEV of $T$), while in other cases it can be as small as a loop factor (as is familiar for saxion couplings). 

Decays of moduli through other operators are discussed in, for example, Refs.~\cite{Moroi:1999zb,Kaplan:2006vm,Endo:2006zj,Nakamura:2006uc,Asaka:2006bv,Dine:2006ii,Endo:2006tf,Acharya:2008bk,Bose:2013fqa}. Given a particular supergravity model, one could use their results to determine the constant $c$ in Eq.~\ref{eq:moduliwidth}. Remaining agnostic about models, we will consider a range of values,
\beq
10^{-3} < c < 100, 
\label{eq:crange}
\eeq
with the low end being ``saxion-like'' (in the sense of potentially arising from $\lambda \sim \alpha_s/\pi$) and the high end resulting from order-one coefficients and a large number of open decay channels. (For a particular model, Ref.~\cite{Acharya:2008bk} computes the coefficient and finds $c = 8\pi$ and $16\pi$ for two different moduli fields, more consistent with the upper end of our range of $c$ values.) 

\subsection{Constraints on the reheating temperature}

\begin{figure}[!h]\begin{center}
\includegraphics[width=0.45\textwidth]{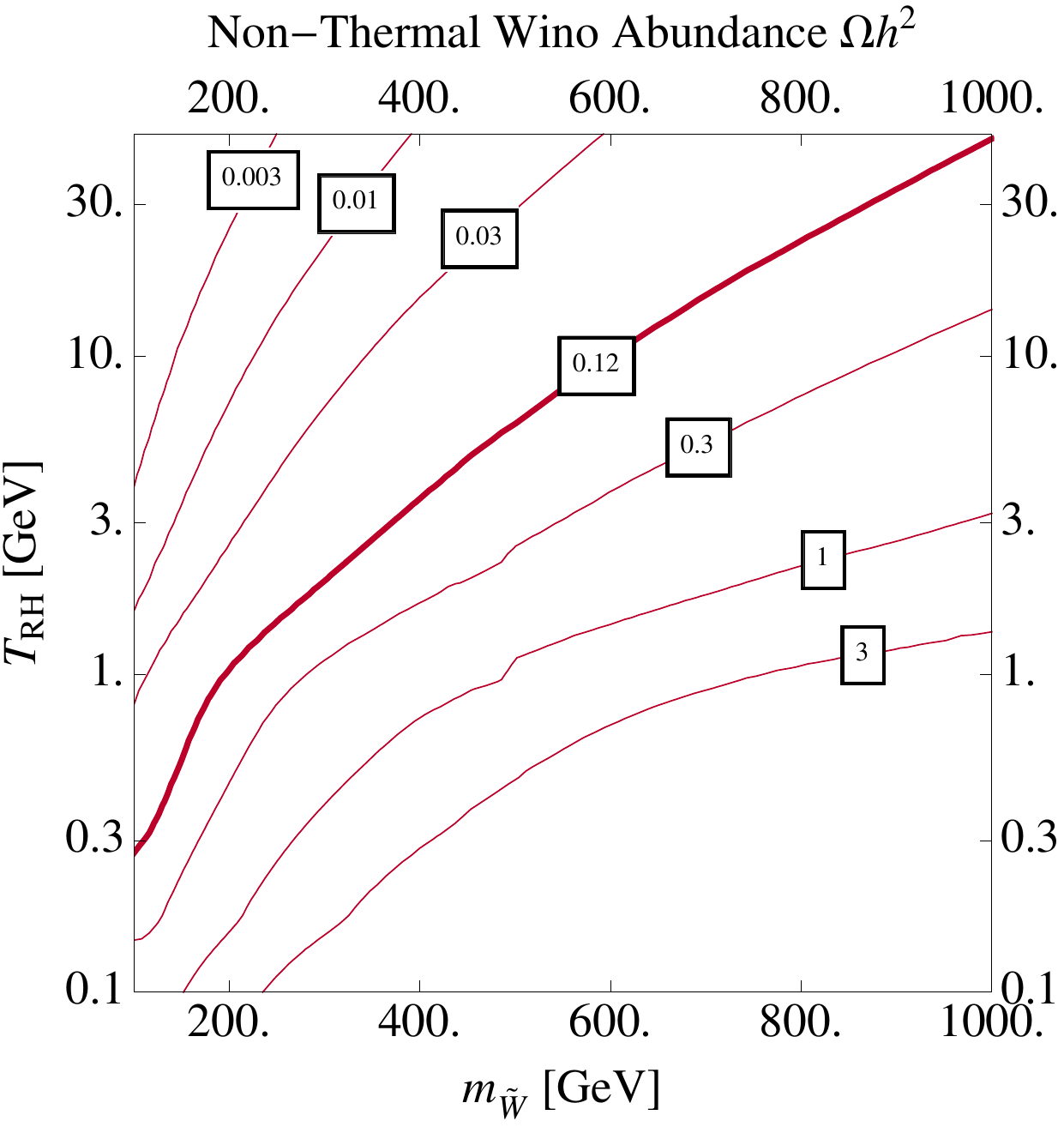}
\end{center} 
\caption{The wino relic density in a non-thermal cosmology with decaying moduli fields, as a function of the wino mass and reheating temperature. The reheating temperature is chosen by convention to be related to the modulus decay width as in Eq.~\ref{eq:Treheatdef}.}
\label{fig:winorelicabundance}
\end{figure}

The results of Sec.~\ref{sec:con} have given a bound (or range of bounds, depending on astrophysical assumptions) on $\Omega_{\tilde W} h^2$ as a function of the wino mass. We would like to translate this into a bound on the moduli mass; up to a choice of the factor $c$ in the decay width, this is equivalent to a bound on the width $\Gamma_\phi$. We compute $\Omega_{\tilde W} h^2$ as a function of $\Gamma_\phi$ by solving the Boltzmann equations for the energy density in moduli particles ($m_\phi n_\phi$), winos ($m_{\tilde W} n_{\tilde W}$), and radiation ($\rho_{\rm rad}$). We choose initial conditions at temperatures large compared to $T_{\rm RH}$, at which time the energy density is dominated by the mass of moduli fields, with a subdominant radiation component of temperature $T \sim (HT_{\rm RH}^2 M_{\rm Pl})^{1/4}$ from moduli that have already decayed~\cite{Chung:1998rq,Giudice:2000ex}. The relevant equations and their approximate solution have been discussed in the literature, e.g.~\cite{Moroi:1999zb,Arcadi:2011ev,Moroi:2013sla}. An approximate analytic formula is  
\beq
\Omega_{\tilde{W}}^{\rm non-thermal} h^2 \approx \frac{3 \Gamma_\phi}{\langle \sigma_{\rm eff} v \rangle s(T_{\rm RH})}\frac{m_{\tilde W}}{3.6 \times 10^{-9}~{\rm GeV}},
\label{eq:case2}
\eeq
where $s(T_{\rm RH}) = 2\pi^2 g_{*s}(T_{\rm RH})T_{\rm RH}^3/45$ is the entropy density at temperature $T_{\rm RH}$. Notice that this is independent of the number of winos produced per moduli decay.\footnote{This is always true as long as the number of winos produced per late-decaying particle decay is $> 10^{-4}$, which is generically true for moduli models as explained in the previous section.} Roughly speaking, the wino relic abundance is enhanced relative to the thermal case by a factor of $T_{\rm freezeout}/T_{\rm RH}$, but in detail this statement is oversimplified (for instance, due to the differing values of $g_*(T)$ at the freezeout and reheat temperatures). In solving the Boltzmann equations, we include several effects that potentially make order-one differences in the result. Because $g_*(T)$ and $g_{*s}(T)$ vary rapidly near the QCD scale, we compute them from a table in the DarkSUSY code~\cite{Gondolo:2004sc}, and we do not omit factors involving $\partial\log g_{*(s)}/\partial\log T$ in the Boltzmann equation for $\rho_{\rm rad}$. Furthermore, because the Sommerfeld effect becomes important for heavy winos~\cite{Hisano:2006nn,Hryczuk:2011vi,Moroi:2013sla}, we have computed the temperature-dependent value of $\left<\sigma_{\rm eff} v\right>$ from a preliminary version 1.1 of the DarkSE code~\cite{Hryczuk:2011tq}, kindly provided to us by Andrzej Hryczuk, taking into account not only the Sommerfeld effect but also coannihilation among the different wino species. As an input to this code, we use the two-loop splitting between the neutral and charged winos from Ref.~\cite{Ibe:2012sx}. For wino masses of about a TeV and temperatures around a GeV, the Sommerfeld enhancement can be as large as a factor of 3 in $\left<\sigma_{\rm eff} v\right>$. We show our computed contours of constant $\Omega_{\tilde W} h^2$ in the plane of wino mass and reheating temperature in Figure~\ref{fig:winorelicabundance}. To the extent that our numerical results differ slightly from the recent results of Ref.~\cite{Moroi:2013sla}, we believe it is mainly due to our more elaborate treatment of the functions $g_*(T)$ and $g_{*s}(T)$.

\begin{figure}[!h]\begin{center}
\includegraphics[width=0.7\textwidth]{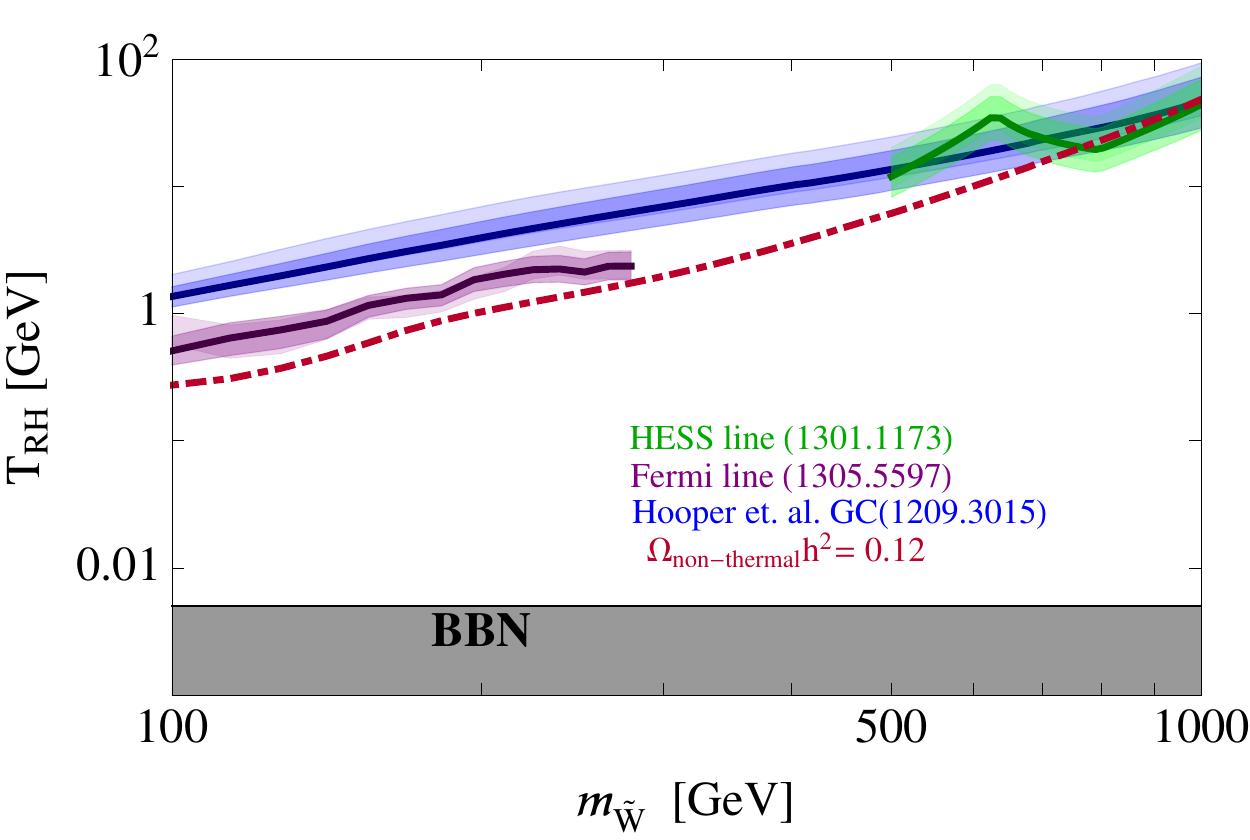}
\end{center} 
\caption{Lower bounds on modulus reheating temperature $T_{RH}$ as a function of wino mass. The blue, purple, green curves with bands around them correspond to constraints from Fermi galactic center continuum, Fermi line search and HESS line search respectively. $T_{RH}$ has to be above 5 MeV (the black solid line) for a successful BBN. The burgundy dot-dashed line is the curve when $\Omega_{\rm non-thermal} h^2= 0.12$. }
\label{fig:reheatingbig}
\end{figure}

Having determined $\Omega_{\tilde W}h^2$ as a function of $T_{\rm RH}$, we can then invert the relationship and express our bound on $\Omega_{\tilde W} h^2$ as a lower bound on the reheating temperature. We have plotted this bound in Figure~\ref{fig:reheatingbig}.

\subsection{Implications for the SUSY breaking scale and cosmology}

\begin{figure}[!h]\begin{center}
\includegraphics[width=0.7\textwidth]{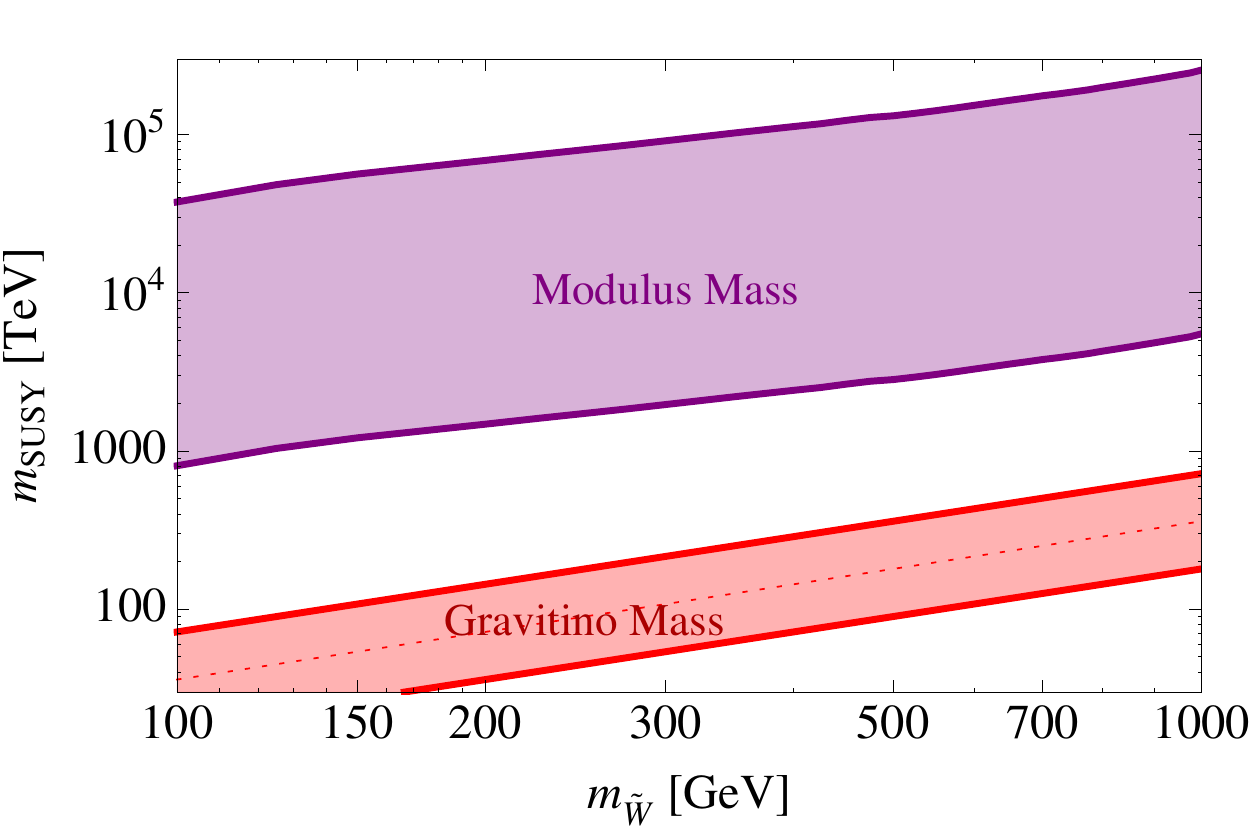}
\end{center} 
\caption{The bound on reheating temperature converted to a lower bound on the scale of moduli masses using Eq.~\ref{eq:Treheatdef} and~\ref{eq:moduliwidth}. Rather than using the reheating bound from Fig.~\ref{fig:reheatingbig}, we have been somewhat more conservative by using the bound on $\left<\sigma v\right>_{WW}$ assuming a 1 kpc cored NFW profile from Fig.~\ref{fig:core} (right hand plot). We also show a range of gravitino masses that might be associated with a given wino temperature. The central dashed line is the AMSB prediction, and the band encompasses a factor of 2 around this prediction in either direction.}
\label{fig:modulusmass}
\end{figure}

We have now bounded the reheating temperature, and we would like to translate this to a statement about the SUSY breaking scale. In any given model of moduli stabilization and decay, this could be translated directly into a statement about the supersymmetry breaking scale. It is hard to do this in complete generality. However, the mini-split SUSY scenario we consider is arguably most natural when there is a loop factor separating the gravitino mass and the gaugino masses, as predicted for example by anomaly mediation~\cite{Randall:1998uk,Giudice:1998xp} but also by other constructions like many IIB flux compactifications~\cite{Choi:2005ge,Choi:2005uz,Conlon:2006us,Conlon:2006wz} or the $G_2$-MSSM~\cite{Acharya:2007rc,Acharya:2008zi}. In the case of anomaly mediation, the prediction is $M_2 = \beta(g)/g m_{3/2} \approx m_{3/2}/360$~\cite{Wells:2003tf}. We plot this as a dashed red line in Fig.~\ref{fig:modulusmass}, and also plot a band that is a factor of two around this prediction, which could be thought of as representing a range of plausible outcomes in other models where the detailed numerical coefficient is sensitive to moduli stabilization or other dynamics. What is clearly visible in Fig.~\ref{fig:modulusmass} is that {\bf the moduli mass scale preferred for achieving a sufficiently small wino relic abundance is notably larger than the gravitino mass expected to lead to the chosen wino mass}. 

To restate this: scenarios in which gauginos are a loop factor below $m_{3/2}$ and moduli lie near $m_{3/2}$ are disfavored, whereas moduli an order of magnitude or more heavier than $m_{3/2}$ are compatible with the data. The modulus mass can only be significantly heavier than the gravitino mass if moduli are stabilized in a supersymmetric manner. Furthermore, it would be a surprise if {\em all} moduli are stabilized supersymmetrically. For instance, if a QCD axion originates from a modulus field, its scalar superpartner, the saxion, would be catastrophically light unless it is stabilized in a nonsupersymmetric manner~\cite{Banks:2002sd,Conlon:2006tq,Bose:2013fqa}. Hence, we might expect the saxion to overproduce winos. This may not be an insurmountable problem: if the axion's decay constant is relatively small, perhaps the saxion stored a small fraction of the energy density compared to other moduli, and hence is a subdominant effect compared to heavier, supersymmetrically stabilized moduli. Another possible problem is that moduli heavy relative to $m_{3/2}$ will decay to gravitinos, potentially creating a moduli-induced gravitino problem~\cite{Kaplan:2006vm,Endo:2006zj,Nakamura:2006uc}. The decay rate of gravitinos in the MSSM is~\cite{Nakamura:2006uc}
\beq
\Gamma_{3/2} = \frac{193}{384\pi} \frac{m_{3/2}^3}{M_{\rm Pl}^2},
\eeq
parametrically similar to moduli decay with $c\approx 2$ but with $m_\phi$ traded for the smaller $m_{3/2}$. This can be problematic for BBN; for instance, 100 TeV gravitinos decay when the temperature is about 7.8 MeV. The gravitino decays also produce additional LSPs, which at these later times do not annihilate as efficiently. As a result, the data appears to be forcing us into a special corner of model space in which moduli decays to gravitinos are suppressed~\cite{Dine:2006ii,Bose:2013fqa}. This problem, known for several years, is more severe now that data has told us that low-mass winos can constitute at most a small fraction of the dark matter. The bound on the reheating temperature is such that we can't appeal to moduli lighter than $2m_{3/2}$ to escape the problem, at least unless gaugino masses are suppressed far below their anomaly-mediated values relative to $m_{3/2}$.

\section{Other possible neutralino dark matter scenarios}
\subsection{Pure bino dark matter}
\label{sec:binohistory} 
For pure bino dark matter, in (mini-)split SUSY or any other scenario without coannihilation, the thermal relic abundance is always many orders of magnitude above the observed dark matter abundance. To lower it to the right relic abundance, a huge late-time entropy dilution is required, for instance from a late-decaying particle that has a tiny branching fraction into binos. Toy models to realize this have been demonstrated in~\cite{Chung:1998rq, Giudice:2000ex}, where the number of dark matter particles produced per decay has to be tiny, e.g., $\lesssim 10^{-7}$. In this case, the dominant mechanism for producing binos is not directly from decays but rather from particles in the SM plasma that annihilate into bino pairs, and the final abundance can be {\em proportional} to $\left<\sigma v\right>$, rather than inversely proportional to it as in standard thermal scenarios. However, generically moduli that decay to SM fields have an order one branching fraction into neutralinos, so this scenario does not appear likely in conventional models of moduli domination and decay.

\subsection{Pure higgsino dark matter and mixed higgsino/bino dark matter} 
\label{sec:higgsinohistory}

Higgsinos, like winos, occur in a multiplet with nearly degenerate states, but unlike winos the different higgsino states are split by a dimension 5 operator and so the splitting will tend to be on the order of a GeV or larger (as explained in more detail in Appendix~\ref{app:masssplitting}). Higgsino dark matter requires a small $\mu$ parameter. In mini-split SUSY, the criterion of an approximately vanishing eigenvalue of the Higgs mass matrix leads to the relation
\beq
\tan\beta = \sqrt{\frac{m_{H_d}^2 + \left|\mu\right|^2}{m_{H_u}^2+\left|\mu\right|^2}}.
\eeq
Because of this, we expect a close association between large values of $\mu$ and $\tan \beta$ values near 1 (since there is no particular reason to expect $m_{H_d}^2 \approx m_{H_u}^2$). Higgsino dark matter is associated with smaller values of $\mu$, and hence somewhat larger values of $\tan\beta$. Because the scale of scalar masses required for a 125 GeV Higgs is strongly $\tan\beta$-dependent, this means that the upper end of the scalar masses considered in mini-split SUSY is unlikely to occur with higgsino dark matter. As one can see from Fig. 2 in~\cite{ArkaniHamed:2012gw}, for $\tan\beta>1$, the SUSY scale has to be below about $10^3$ TeV. For $\tan\beta >4$, the scale has to be even lower, below 100 TeV. Thus, a discovery of higgsino dark matter could be interesting to help narrow the range of plausible scalar masses to target in future experiments.

Because higgsinos have smaller annihilation cross sections than winos, they tend to be overproduced in nonthermal scenarios with low reheating temperatures. Thus, they are less favored as dark matter candidates in scenarios with decaying moduli, although still viable with relatively high reheating temperatures. They could also be  a component of dark matter in the multi-component dark matter scenario~\cite{Baer:2011hx, Bae:2013qr, Graf:2013xpe}. 
On the other hand, mixed bino/higgsino dark matter can provide a thermal dark matter candidate over a range of masses, provided $M_1$ and $\mu$ are tuned to be similar, a case known as the well-tempered neutralino~\cite{ArkaniHamed:2006mb}.

\begin{figure}[!h]\begin{center}
\includegraphics[width=0.8\textwidth]{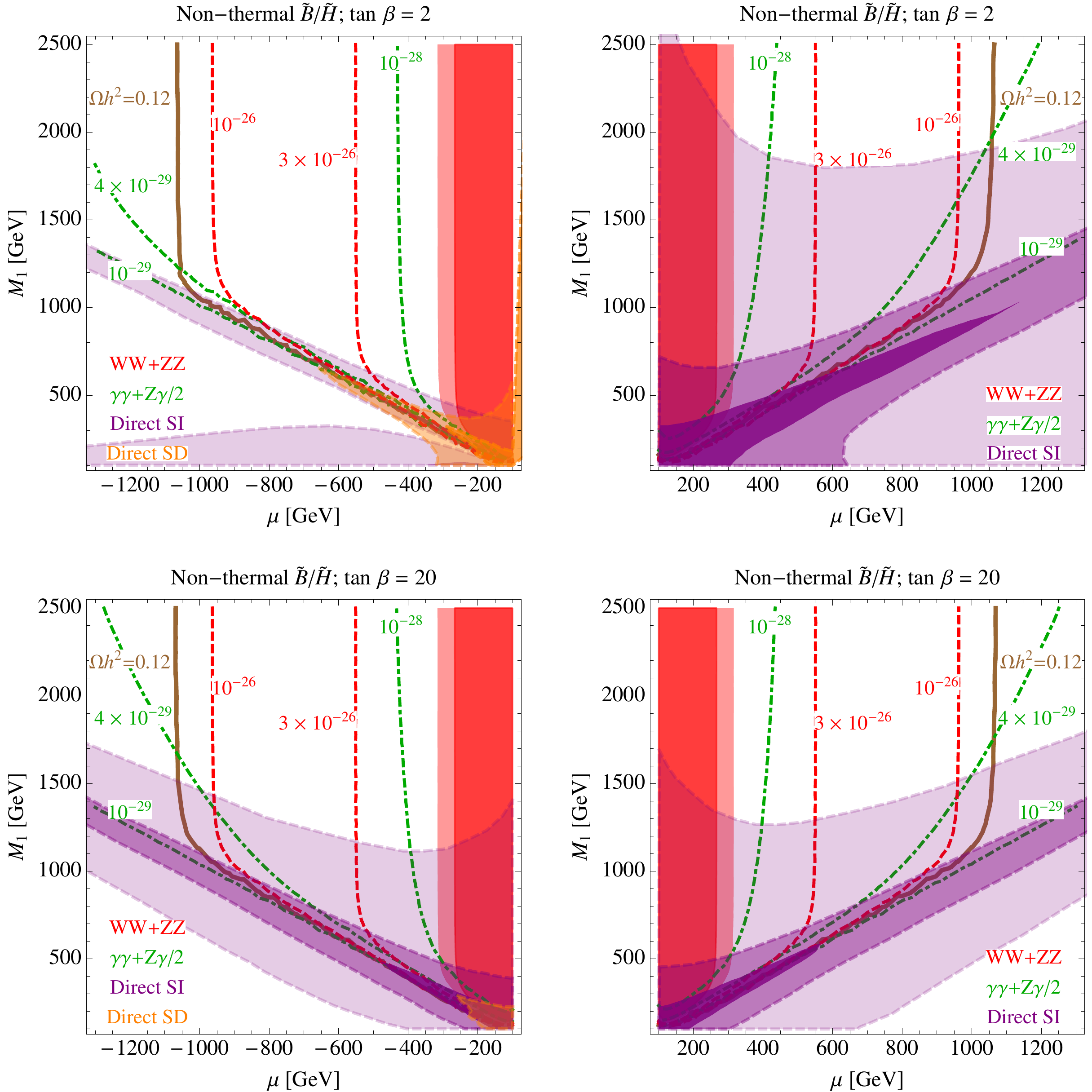}
\end{center}
\caption{Complementarity of direct and indirect detection in the higgsino/bino plane. The direct detection bounds are extracted from Ref.~\cite{Cheung:2012qy}. The darkest purple region is the current XENON100 bound on spin-independent dark matter--nucleus scattering. The two surrounding lighter purple regions are the projected LUX and XENON1T bounds, respectively. The dark orange shaded region in the top left plot, and that in the bottom left plot, are IceCube bounds on the spin-dependent dark matter--nucleus scattering rate (assuming annihilation to $W^+W^-$) while the lighter orange region in the top left plot is the XENON1T spin-dependent projected reach. The red shaded regions are Einasto (lighter) and NFW (darker) exclusions from Ref.~\cite{Hooper:2012sr}. Dot-dashed green curves show gamma-ray line rates and dashed red lines show gamma-ray continuum rates, computed with MicrOMEGAs~\cite{Belanger:2013oya}.}
\label{fig:binohiggsinoplane}
\end{figure}%

Mixed higgsino/wino dark matter will tend to interpolate between pure higgsinos and the pure wino case we have already discussed, with the wino component allowing higgsinos to be more easily identified in indirect detection. The case of mixed higgsino/bino dark matter deserves a longer discussion. Because binos produce no indirect detection signal (at least in mini-split scenarios), adding a bino admixture to higgsinos reduces the prospects for signals in gamma rays. However, the existence of a bino--higgsino--Higgs vertex allows mixed bino/higgsino states to scatter with nuclei through Higgs exchange, leading to bounds from {\em direct} detection experiments. The prospects for complementarity of indirect and direct detection have been noted recently in~\cite{Bergstrom:2010gh,Cahill-Rowley:2013dpa}. Recent discussions of the bounds on mixed bino/higgsino dark matter have appeared in Refs.~\cite{Perelstein:2012qg,Cheung:2012qy}. The lesson is that pure higgsinos are unconstrained, but even a relatively small bino mixture can run into conflicts with XENON100 bounds. However, the detailed bounds depend on the relative sign of $\mu$ and $M_1$ as well as on $\tan\beta$, and certain direct detection ``blind spots'' exist.

We have scanned the mixed bino/higgsino parameter space and computed the cross section for indirect detection signals in gamma rays. We use MicrOMEGAs to compute the relic abundance~\cite{Belanger:2001fz,Belanger:2004yn,Belanger:2013oya}.\footnote{In MicrOMEGAs,``Omega = '' gives the value of $\Omega h^2$~\cite{Belanger:2001fz}.} We plot various current and future constraints on mixed higgsino/bino dark matter in Fig.~\ref{fig:binohiggsinoplane}. The direct detection current and prospective bounds (XENON and LUX spin-independent, XENON spin-dependent, and IceCube from solar capture followed by annihilation to $W^+W^-$) are all extracted from figures in Ref.~\cite{Cheung:2012qy}. We have superimposed the Fermi-LAT gamma ray continuum limit from Ref.~\cite{Hooper:2012sr}. The gamma-ray continuum and line rates were computed with MicrOMEGAs, which includes diagrams relevant for higgsino annihilation to $Z\gamma$ that are absent from the early literature and treated for the first time in Ref.~\cite{Boudjema:2005hb}.

The figure displays an important complementarity between direct and indirect detection. As already noted, direct detection bounds arise dominantly from Higgs exchange, which depends on the Higgs--bino--higgsino coupling and hence on having $\mu \sim M_1$. The large bino mixture leads to a larger splitting between charged and neutral states and decreases indirect detection rates. In the pure higgsino limit, on the other hand, indirect detection both from continuum annihilation to $W^+W^-$ and $ZZ$ and from gamma-ray lines has much better reach. In particular, {\em either} ruling out continuum gamma rays from $\left<\sigma v\right>_{WW+ZZ} \approx 10^{-26}~{\rm cm}^3/{\rm s}$ or gamma-ray lines with $\left<\sigma v\right>_{\gamma\gamma+Z\gamma/2} \approx 4 \times 10^{-29}~{\rm cm}^3/{\rm s}$ would, together with bounds from XENON, essentially completely rule out (or discover!) mixed bino/higgsino dark matter. Given the reach quoted in a recent look at the prospects for future gamma-ray observatories to detect lines~\cite{Bergstrom:2012vd}, it appears that HESS-II, CTA, and GAMMA-400 will all fall significantly short of the reach needed to exclude thermal pure higgsino dark matter, although they could still set interesting constraints on lighter nonthermal higgsinos. There are brighter near-future prospects for progress from continuum photons; CTA, for example, expects to probe this rate at a level of few times $10^{-26}~{\rm cm}^3/{\rm s}$~\cite{Doro:2012xx}, which would eliminate a very large fraction of the remaining mixed bino/higgsino parameter space. (For more on how CTA will constrain neutralino parameter space, see the recent Ref.~\cite{Cahill-Rowley:2013dpa}.) It appears that thermal higgsinos will still elude the reach of near-future experiments, but they provide a very important benchmark to aim for in future observations.

\section{Discussion}

\subsection{Wino and Higgsino Benchmarks for Indirect Detection}

The progress of indirect detection experiments is often measured by the bounds they can place on light WIMPs annihilating to final states like $b{\bar b}$ and $\tau^+\tau^-$. But some of the most natural candidates for WIMPs come from simple SU(2)$_L$ multiplets, with winos and higgsinos serving as important examples~\cite{Cirelli:2005uq,Cirelli:2007xd}. Such WIMPs have distinctive features: coannihilation among different states in the multiplet and Sommerfeld-enhanced cross sections are important for determining their relic abundance, and their loop-level annihilation rates leading to gamma ray line signatures can be surprisingly large at high mass. These multiplets are important benchmarks for indirect detection experiments to target, especially given their strong motivation from models of supersymmetry.

In fact, as we have argued in Section~\ref{sec:constraints}, the current generation of gamma ray observatories---especially Fermi-LAT and HESS---are approaching a major milestone that has gone almost unheralded. It is already the case that a combination of their data excludes wino dark matter over the entire range of interesting masses, {\em assuming} a favorable halo profile. This is a strong assumption, as the Milky Way's dark matter halo is relatively unconstrained, but it brings the complete exclusion of wino dark matter tantalizingly within reach. Future observations will push the limits down, ameliorating the effect of astrophysical uncertainties. Strategies like searching for gamma rays from locations somewhat off the galactic center may also help to reduce the impact of uncertain halo profiles.  But it is also important to reduce the astrophysical uncertainties. The exclusion of thermal relic winos currently hinges on the answer to an important question about the physics of galaxies: can baryonic feedback effects, such as bar--halo interactions, produce a core of approximately kiloparsec size in a spiral galaxy like the Milky Way? Such questions are intrinsically interesting, but take on even more importance when they are so strongly linked to crucial questions in particle physics.

Even short of an answer, a better quantification of our uncertainty could be useful. For instance, a reasonable prior distribution on a suitable space of halo shapes, which one could marginalize over when setting limits, could allow us to make a clearer statement about the extent to which current bounds depend on uncertain astrophysics. Such a prior would have to be chosen with care by experts in the possible range of baryonic feedback effects.

Given the astrophysical uncertainties, the story of pure wino dark matter is not yet complete. But it is a beautiful example. It illustrates the power and reach of the current generation of cosmic ray experiments like Fermi-LAT and HESS, which have now constrained a well-motivated theoretical dark matter candidate over essentially the entire range of viable masses. The case of pure higgsino dark matter, with its associated smaller cross sections, provides a further benchmark to target with future observations. This is a problem nicely located at the intersection of theoretical particle physics, experimental cosmic ray physics, and the dynamics of galaxy formation, and refinements on all three fronts can help to extend and complete the story. 

\subsection{The Case for a Low Reheating Temperature}

As we have discussed in Section~\ref{sec:moduli}, the strong bounds on light wino dark matter below about 200 GeV---which is ruled out by a large factor even when assuming a flattened NFW profile with a 1 kiloparsec core---can be interpreted as very strong constraints on nonthermal cosmologies in which moduli decay at late times. Here we will briefly make a case for why such a cosmology is well-motivated and important, before turning to a look at what scenarios survive the bounds.

The observed Higgs boson mass of 125 GeV requires, in the MSSM, heavy stops (possibly with large $A$-terms), which inevitably requires a high degree of fine-tuning. This is suggestive of a split spectrum, with scalar superpartners heavy relative to the gauginos (except for a single Higgs doublet, tuned to be light)~\cite{Hall:2011jd,Arvanitaki:2012ps,ArkaniHamed:2012gw,Hall:2012zp}. One of the most well-motivated versions of this scenario has gauginos a loop factor below the gravitino mass $m_{3/2}$ and scalar masses at the scale $m_{3/2}$. This is the prediction of anomaly mediation~\cite{Giudice:1998xp}, unless there is a special ``sequestering'' mechanism~\cite{Randall:1998uk}. In supergravity models motivated by string compactifications, there are typically moduli-mediated contributions to soft masses. In a wide variety of constructions, these lead to tree-level contributions to gaugino masses that are parametrically smaller than $m_{3/2}$ by factors like $\log\left(M_{\rm Pl}/m_{3/2}\right)$~\cite{Choi:2005ge,Choi:2005uz,Conlon:2006us,Conlon:2006wz,Acharya:2007rc,Acharya:2008zi}. In some of these constructions, scalar masses are claimed to be of the same order, but this depends on model-dependent details of how they couple to moduli, and in some cases the scalar masses are of order $m_{3/2}$~\cite{Acharya:2007rc,Acharya:2008zi}. Some of the tree-level factors in gaugino masses are loop factors in disguise. For instance, a simple model of a modulus $T$ with a constant superpotential plus a gaugino condensation term $A e^{-bT}$ predicts~\cite{Chacko:2000fn}
\beq
m_\lambda = \frac{F_T}{T} = \frac{2}{bT} m_{3/2} = \frac{g^2 N}{4\pi^2}m_{3/2},
\eeq
since $bT$ is precisely the fractional instanton action $8\pi^2/(g^2 N)$. Quite generally, the factors like $\log(M_{\rm Pl}/m_{3/2})$ and other ``tree-level'' numerical suppression factors of gaugino masses in moduli mediation scenarios turn out to be equivalent to loop factors once the moduli VEVs appearing in these expressions are related to gauge couplings.

Even if it were a robust prediction of supersymmetry breaking models that gauginos are a loop factor below scalars in mass, it isn't clear why this is necessarily linked to fine-tuning. We could have lived in a world that was split {\em and} natural, with all the scalars at around 100 GeV to 1 TeV and gauginos a loop factor lighter. A world with light gauginos is not obviously a world incompatible with our existence; it is only because of direct collider constraints on light gauginos that we know that they don't exist. One possible explanation, however, comes from the moduli problem. Assuming that moduli fields with mass of order the scalar soft masses exist, the natural split SUSY scenario would predict that moduli would dominate the universe until late times. Moduli masses at 100 GeV lead to reheating temperatures ranging from about 7 eV up to 2 keV (for widths corresponding to Eqs.~\ref{eq:moduliwidth},\ref{eq:crange}). In such a universe, BBN would be very different. The faster expansion rate of the universe would lead to weak interactions freezing out at higher temperatures, and hence equal numbers of neutrons and protons at the time that BBN begins. The temperatures at which nuclei form would also be reached at earlier times, so very few of the neutrons would have decayed. As a result, we expect that baryons would have assembled themselves mostly into helium nuclei. It is very plausible that life could not exist in such a universe, where stars as we know them would not burn and the cooling processes that drive structure formation would be radically different. It is not completely clear that this anthropic argument is strong enough to exclude larger reheating temperatures, around 0.1 MeV, although even in this case we expect the hydrogen abundance would be significantly suppressed compared to our universe.\footnote{We thank Josh Ruderman for discussions that helped to clarify this idea.}

This leads to an appealing conclusion: although the Higgs mass and LHC data so far offer bleak prospects for natural SUSY, it could be that a universe with truly natural SUSY is a universe that would be inhospitable to life. Within the framework of moduli cosmology, it may be that we live in the {\em most natural universe that allows us to exist}. Such a universe would have SUSY split by a loop factor, with gauginos at the weak scale, moduli near 30 TeV, and a nonthermal cosmology that reheated the universe to just above the temperature of BBN. This scenario for SUSY breaking and cosmology has been advocated in Refs.~\cite{Kaplan:2006vm,Acharya:2008zi,Acharya:2009zt}, although to the best of our knowledge the anthropic constraint from avoiding an all-helium universe has not previously been discussed, and so existing literature has left open the question of why we have not found ourselves in a natural split SUSY universe. We find the resulting ``minimally tuned anthropic universe'' plausible and interesting. More detailed work on the cosmological history and anthropic constraints are needed to assess how sharp the argument can be, but it seems to robustly exclude the possibility of a fully natural split SUSY spectrum.

\subsection{SUSY Scenarios in Light of the Bounds}

The bounds on light wino dark matter that we have discussed pose a problem for the scenario in which the wino relic abundance is set by moduli decays. In fact, even before applying the bounds, the reheating temperature that corresponds to a wino relic abundance of $\Omega h^2 = 0.12$ is already above 200 MeV, as shown in Fig.~\ref{fig:winorelicabundance}, which is far enough above the temperature of BBN to undermine any anthropic argument. The heavy mass of the moduli needed for this reheating temperature was also emphasized recently in Ref.~\cite{Bose:2013fqa}. Furthermore, as discussed in Section~\ref{sec:moduli}, indirect detection data requires an even smaller value of $\Omega h^2$, demanding very heavy moduli. Such heavy moduli are typically associated with a severe moduli-induced gravitino problem. This puts the moduli-dominated cosmology of Refs.~\cite{Moroi:1999zb,Kaplan:2006vm,Acharya:2008zi,Acharya:2009zt} in a precarious position.

However, we would like to suggest a way to salvage the picture of a low reheating temperature and a minimally tuned anthropic universe. It is possible that the moduli have masses at around $m_{3/2}$, light enough to avoid creating a gravitino problem, and that they reheat the universe to temperatures of only a few MeV. The troubling overabundance of winos can be eliminated by the simple expedient of turning on $R$-parity violation, so that the winos decay. The lifetime of winos could still be long on collider scales, and thus disappearing track searches could still be important for this case.

This scenario would still have a dark matter candidate motivated strongly by top-down considerations: the axion. String constructions often have axions with string-scale decay constants~\cite{Svrcek:2006yi}. In conventional (radiation-dominated) cosmologies, such axions overclose the universe. However, in scenarios with a low reheating temperature just above the bound from BBN, axions with decay constants up to about 10$^{15}$ GeV are viable~\cite{Kawasaki:1995vt,Banks:2002sd}. In this scenario, detection of axion dark matter might be achieved through measurements of the rapidly oscillating neutron EDM it induces~\cite{Graham:2011qk,Graham:2013gfa,Budker:2013hfa}. Another signature of the nonthermal history could arise through effects on inflationary observables like the spectral index~\cite{Easther:2013nga}.

These are not the only options. One could consider scenarios in which moduli are simply decoupled, although we expect this comes with a large tuning cost. Or one could consider low-scale SUSY breaking with low-scale inflation to dilute the abundance of light moduli~\cite{Randall:1994fr}. The required dilution is especially large due to X-ray bounds on decaying moduli~\cite{Kawasaki:1997ah}, and we have found that building a full model consistent with the bounds is difficult, and may lead to strong restrictions on the SUSY-breaking scale~\cite{Fan:2011ua}.

It is an exciting time for the study of supersymmetry. The LHC's discovery that the Higgs mass is 125 GeV heightened the tension between the beautiful idea of SUSY naturalness and the harsh reality of data. The lack of any signature of superpartners at the LHC so far has only made the problem starker. Now, as we have seen, gamma rays from the center of the galaxy have also sharply constrained some of the most compelling remaining models of supersymmetry. Pure higgsino dark matter provides a concrete target for future gamma ray observations. An $R$-parity violating decay of a wino LSP provides a challenging target for collider experiments. Both terrestrial experiment and astronomical observation will be necessary in the rapidly progressing quest to corner weak-scale supersymmetry.

\section*{Acknowledgments}
We learned of the HESS line search indirectly through word of Nima Arkani-Hamed's talk at the winter Aspen conference. We thank Mariangela Lisanti for informing us of her related work in Ref.~\cite{Cohennewpaper}. We thank Andrea Albert, Patrick Draper, Christian Farnier, Gordy Kane, Ian Low, David Pinner, Joshua T. Ruderman, Scott Watson, Neal Weiner and Christoph Weniger for useful discussions and correspondence. We especially thank Andrzej Hryczuk for explaining his Sommerfeld enhancement calculation to us and providing a prerelease version 1.1 of his DarkSE code for our use. JF and MR are supported in part by the Fundamental Laws Initiative of the Harvard Center for the Fundamental Laws of Nature. JF and MR thank the KITP in Santa Barbara for its hospitality while a portion of this work was completed. This work at the KITP was supported in part by the National Science Foundation under Grant No. NSF PHY11-25915.

\appendix
\section{Mass differences between charginos and neutralinos}
\label{app:masssplitting}
In this appendix, we review the mass differences between charginos and neutralinos for higgsino and wino multiplets. In the pure wino limit, $M_2 \ll \mu, M_1$, the tree-level mass difference between charged and neutral winos is:
\beq
\delta m^{\tilde W} \approx \frac{M_Z^4}{M_1\mu^2}s_W^2c_W^2\sin^22\beta \approx 5 \times 10^{-6}\,{\rm GeV}\, \frac{(1 \,{\rm TeV})^3}{M_1 \mu^2} \frac{100}{\tan^2 \beta},
\label{eq:winosplit}
\eeq
which is suppressed by three powers of the larger mass scale and is tiny. This could be easily understood from an effective operator analysis (see e.g.~\cite{Howe:2012xe}). When one integrates out higgsinos, dimension five operators are generated
\beq
\sim \frac{g_1g_2}{\mu\tan\beta}(H^\dagger \sigma^i H) \tilde{W}^i \tilde{B},
\eeq
where $\sigma^i$ are SU(2)$_L$ generators. It generates a mass mixing between wino and bino. Integrating out the bino, one finds a dimension seven operator leading to the tree-level splitting in Eq.~\ref{eq:winosplit}. Notice that another dimension five operator that is generated, $H^\dagger H \tilde{W}^2$, shifts all of the winos by a common mass instead of generating a mass splitting. At one-loop level, custodial SU(2) symmetry breaking induces a mass splitting of order 0.16 GeV (calculated at two loops in~\cite{Ibe:2012sx}), which is much larger than the tree-level splitting. We assume this loop-induced splitting throughout the paper.

In the pure higgsino limit, $\mu \ll M_1, M_2$, unlike the wino case, the tree-level mass splitting is only suppressed by one power of the larger mass scale $M_1$ or $M_2$,
\beq
\delta m^{\tilde H} \approx \frac{m_Z^2}{2M_1}c_W^2\left(1-\sin 2\beta\right)+\frac{m_Z^2}{2M_2}s_W^2\left(1+\sin 2\beta\right) \approx  0.5 \, {\rm GeV} \frac{4.5 \, {\rm TeV}}{M_s},
\eeq
where in the second step, we assume for concreteness that $\tan \beta \gg 1$ and $M_2 =2M_1=M_s$. Again, this could be understood easily from an effective operator analysis. Integrating out a heavy bino or wino, one gets dimension five operators such as 
\beq
H_u^\dagger \tilde{H}_uH_d^\dagger \tilde{H}_d,
\eeq
which will lead to a charged/neutral mass splitting after electroweak symmetry breaking. 

\section{Review of wino annihilation rates to gamma-ray lines}
\label{app:winomatching}

\begin{figure}[!h]\begin{center}
\includegraphics[width=0.9\textwidth]{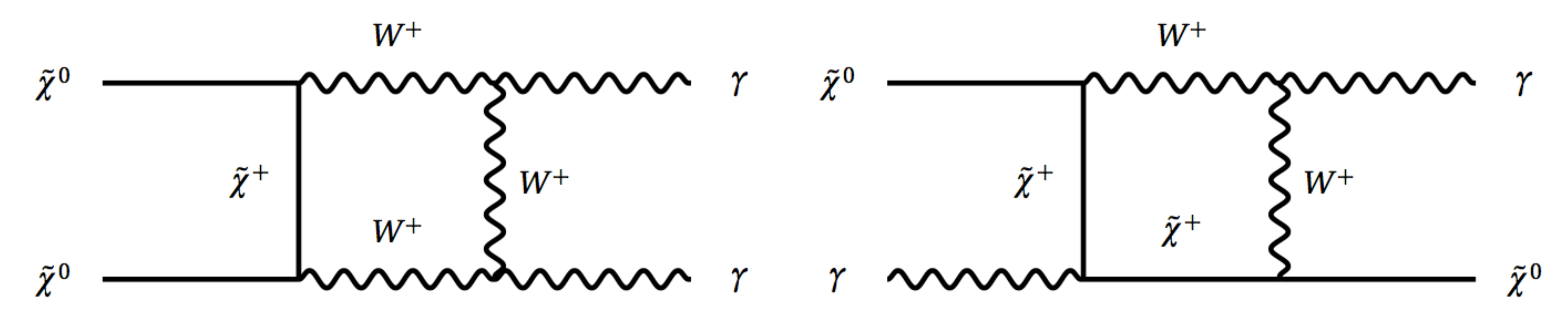}
\end{center}
\caption{Other diagrams (aside from the one shown in Fig.~\ref{fig:diagram}, which dominates for heavy winos) contributing to wino annihilation to photons.}
\label{fig:otherdiagrams}
\end{figure}%

The literature contains several different calculations of wino annihilation to $\gamma\gamma$ and $Z\gamma$~\cite{Bergstrom:1997fh,Ullio:1997ke,Boudjema:2005hb,Bern:1997ng,Hisano:2004ds,Hryczuk:2011vi}. In this appendix, we review some of the salient features of these annihilation rates, and discuss how we collate the various results into a single function to set limits. The first calculation was the one-loop calculation including all diagrams~\cite{Bergstrom:1997fh,Bern:1997ng}. We assume a {\em pure wino}, which eliminates most of the diagrams and leaves the part of the amplitude labeled $\tilde{A}_W$ in Ref.~\cite{Bergstrom:1997fh} as the sole contribution to ${\tilde W}^0 {\tilde W}^0 \to \gamma\gamma$. Inspection of their result shows that it contains pieces with three different denominators:
\beq
\frac{1}{m_{\chi^+}^2 + m_{\chi^0}^2 - m_W^2}, \frac{1}{m_{\chi^+}^2 - m_W^2},~{\rm and}~\frac{1}{m_{\chi^+}^2 - m_{\chi^0}^2 - m_W^2}. \label{eq:denoms}
\eeq
Following Bergstr\"om and Ullio we define:
\beq
a \equiv m_{\chi^0}^2/m_{\chi^+}^2,~{\rm and}~b \equiv m_W^2/m_{\chi^0}^2.
\eeq
For heavy winos, $b$ is a small parameter, while $1 - a$ is always a small parameter because the charged and neutral winos are split dominantly by loops~\cite{Cheng:1998hc,Yamada:2009ve,Ibe:2012sx}, with $\delta m \approx 164$ MeV for heavy winos. For heavy winos, then, the dominant contribution to the annihilation comes from the third denominator in Eq.~\ref{eq:denoms}, which goes as $1/m_W^2$ while the other terms are suppressed by a factor of $b$. The third denominator corresponds to the diagram in Fig~\ref{fig:diagram}, while the other terms are related to other diagrams like those in Fig~\ref{fig:otherdiagrams}. Note that despite the behavior of the denominators, the amplitudes do not blow up when $m_{\chi^+}^2 = m_{\chi^0}^2 + m_W^2$; this is a spurious pole that is canceled due to the identity $I_1(x,1) + 2 I_2(x,1-x) = 0$ (where $I_{1,2}(x,y)$ are defined in Ref.~\cite{Bergstrom:1997fh}). In fact, the $1/m_W^2$ piece of the amplitude is proportional to:
\beq
I_1(a,1) + 2I_2(a,b) \approx 2 \pi \left(\sqrt{1-a} - \sqrt{b}\right),
\eeq 
guaranteeing the necessary cancelation. Putting all the factors together, the leading $1-a$ and $b$ dependence for heavy winos gives:
\beq
\sigma v({\tilde W}^0 {\tilde W}^0 \to \gamma\gamma) \approx \frac{4\pi \alpha^4}{m_W^2 \sin^2\theta_W} \left(1 + \sqrt{\frac{2 \delta m m_{\tilde W}}{m_W^2}}\right)^{-2}.
\label{eq:sigmavapprox}
\eeq
Without the factor in parentheses, the annihilation rate would asymptote to a constant $\approx 1.6 \times 10^{-27}~{\rm cm}^3/{\rm s}$, but because the mass splitting $\delta m$ is approximately constant for heavy winos, the cross section gradually declines with mass.

\begin{figure}[!h]\begin{center}
\includegraphics[width=0.5\textwidth]{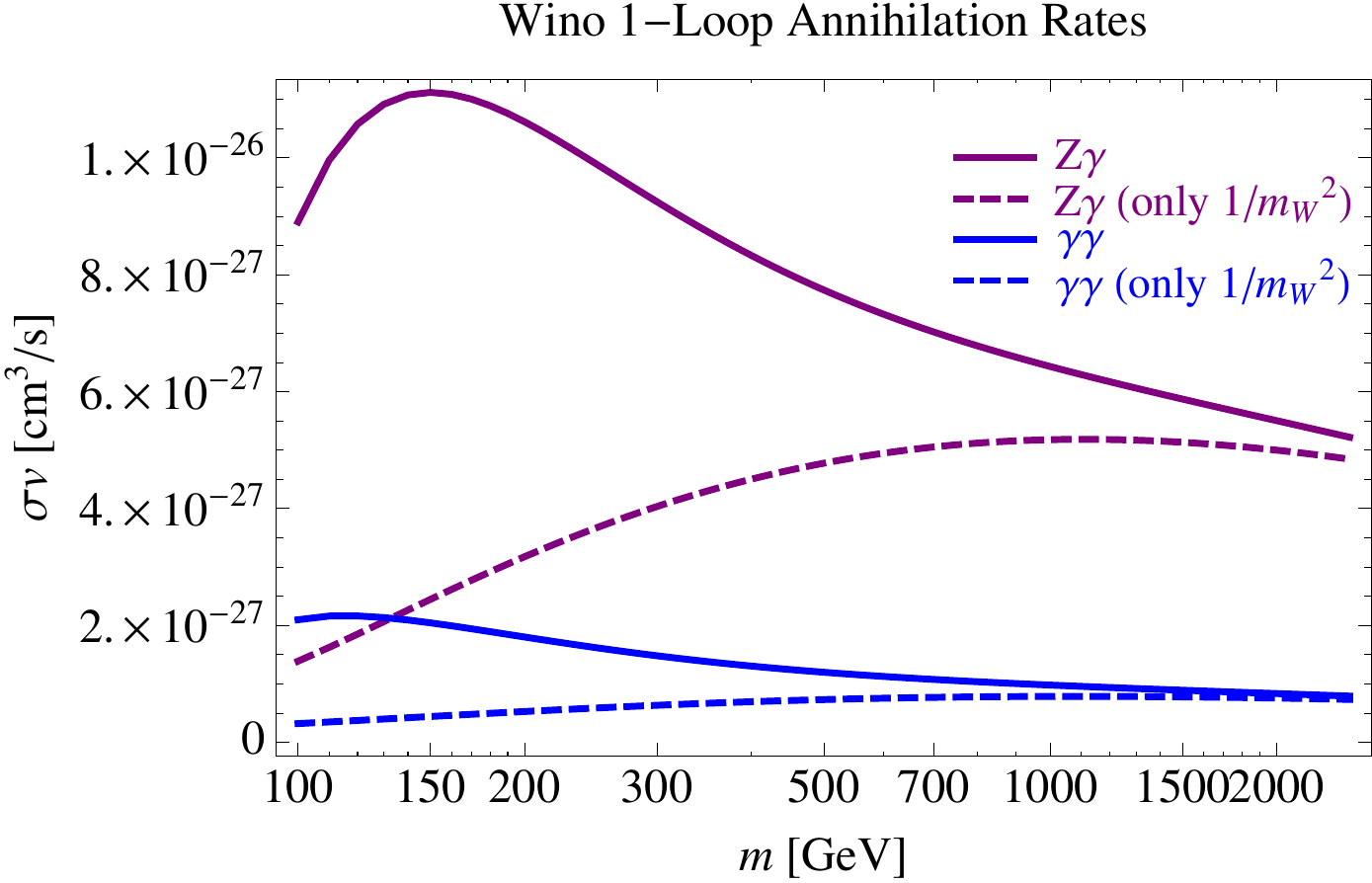}
\end{center}
\caption{One-loop annihilation rates for $\sigma v({\tilde W}^0 {\tilde W}^0 \to \gamma\gamma,Z\gamma)$. We have used the complete one-loop calculations from Refs.~\cite{Bergstrom:1997fh,Ullio:1997ke} and the two-loop charged/neutral wino splitting from Ref.~\cite{Ibe:2012sx}. The dashed lines show the results we would get if we keep only the part of the amplitude proportional to $\left(m_{\chi^+}^2 - m_{\chi^0}^2 - m_W^2\right)^{-1} \approx 1/m_W^2$.} 
\label{fig:winoonelooprates}
\end{figure}%

At large masses, the Sommerfeld effect becomes important, because the weak interactions can be viewed as a long range force for sufficiently heavy winos and can create light bound states when $\alpha m_{\tilde W} \gsim m_W$. Calculations relying on the Sommerfeld effect~\cite{Hisano:2004ds,Hryczuk:2011vi} typically omit the diagrams that do not have $1/m_W^2$ denominators, because such diagrams contribute only to short-range interaction. To assess the mass scale at which this becomes a valid approximation, in Figure~\ref{fig:winoonelooprates} we show both the complete one-loop annihilation rate and the annihilation rate computed when keeping only the third denominator of Eq.~\ref{eq:denoms}.  This makes it clear that at low masses, the diagrams in Fig.~\ref{fig:otherdiagrams} are absolutely crucial to computing the correct rate; keeping only the $1/m_W^2$ piece can underestimate the annihilation rate by a factor of about 5 for winos below 150 GeV. Calculations based on the Sommerfeld effect typically drop these terms, so we have to somehow ``match'' the full one-loop result at small wino masses to the Sommerfeld result at large wino masses.

The simplest approach would be to plot the one-loop result with no Sommerfeld enhancement from Ref.~\cite{Bergstrom:1997fh} and the Sommerfeld-enhanced result of Ref.~\cite{Hryczuk:2011vi} on the same plot and sew the result together where the two curves cross. We have chosen a slightly more subtle matching. The Sommerfeld-enhanced result of Ref.~\cite{Hryczuk:2011vi} also includes subleading one-loop contributions, for instance in the amplitudes for ${\tilde W}^+ {\tilde W}^- \to (Z,\gamma)+\gamma$. These loop effects provide a sizable negative shift in the amplitude already at the scale where the two curves cross, suggesting that the one-loop result of Ref.~\cite{Bergstrom:1997fh} would likely be shifted to substantially smaller values if it were extended to two loops. As a result, we modify the one-loop result by reweighting the part of the amplitude proportional to $\left(m_{\chi^+}^2 - m_{\chi^0}^2 - m_W^2\right)^{-1}$, which is essentially the part that can be interpreted in terms of chargino annihilation, by the correction factor plotted in Fig. 11 of~\cite{Hryczuk:2011vi} for charged wino annihilation. This gives a downward shift in the one-loop result at intermediate masses and leads to matching onto the Sommerfeld-enhanced result at slightly lower mass than we otherwise would. The final result is the solid red curve plotted in Fig.~\ref{fig:line}. This matching procedure is admittedly somewhat ad hoc, but by including a negative correction (when the full two-loop result at low mass is unknown) we expect it to make our result more conservative. It would be an interesting exercise in effective field theory to develop a more sophisticated matching of the standard fixed-order calculations and the Sommerfeld-enhanced result to interpolate between the different regimes.

\section{Non-thermal cosmology with insufficient production of dark matter}
\label{app:non-thermal2}
In this appendix, we discuss another possible non-thermal scenario for wino dark matter where the non-thermal production from a late-decaying particle $\xi$ is not compensated by wino annihilation as opposed to the moduli scenario. The relic abundance of wino dark matter is then estimated to be~\cite{Moroi:1999zb} 
\beq
\Omega_{\tilde{W}}^{\rm non-thermal} h^2\approx \frac{3 b \Gamma_\xi^2 M_{\rm pl}^2}{m_\xi s(T_{RH})}\frac{m_{\tilde{W}}}{3.6 \times 10^{-9}\,{\rm GeV}} \approx 2\times 10^3 \eta\frac{g_*(T_{RH})}{g_{*,s}(T_{RH})} \frac{m_{\tilde{W}}}{100 \,{\rm GeV}} \frac{T_{RH}}{10 \, {\rm MeV}},
\label{eq:case1}
\eeq
where $\eta = b (100\, {\rm TeV}/ m_\xi)$ with $b$ the number of wino per $\xi$ decay. Unlike Eq.~\ref{eq:case2}, the relic abundance in this case scales linearly with reheating temperature and number of wino produced per decay. In this case, unless $\eta < 10^{-4}$ and the reheating temperature is close to the BBN bound 5 MeV, wino will still overclose the Universe.

\bibliography{ref}
\bibliographystyle{utphys}
\end{document}